\documentclass[reprint,amsmath,amssymb,aps,prl,superscriptaddress]{revtex4-2}

\usepackage{graphicx}% Include figure files
\usepackage{dcolumn}% Align table columns on decimal point
\usepackage{color} 
\usepackage{bm}% bold math
\usepackage{hyperref}% add hypertext capabilities
\usepackage{booktabs}% for the tables

\begin{document}

\preprint{APS/123-QED}

\title{What is the diatomic molecule with the largest dipole moment?}% Force line breaks with \\

\author{Ahmed  Elhalawani}
\affiliation{%
Department of Physics and Astronomy, Stony Brook University, 11794 Stony Brook, NY, USA
}%
 \author{Ruiren Shi}
 \affiliation{%
Department of Physics and Astronomy, Stony Brook University, 11794 Stony Brook, NY, USA
}%
 \author{Mateo Londoño}
 \affiliation{%
Department of Physics and Astronomy, Stony Brook University, 11794 Stony Brook, NY, USA
}%
\author{Micha{\l} Tomza}
 \affiliation{%
Faculty of Physics, University of Warsaw, Pasteura 5, 02-093 Warsaw, Poland
}%
\author{Jes\'us P\'erez-R\'ios}%
 \email{jesus.perezrios@stonybrook.edu}
\affiliation{%
Department of Physics and Astronomy, Stony Brook University, 11794 Stony Brook, NY, USA
}%

\date{\today}% It is always \today, today,

\begin{abstract}

We present a machine learning model for predicting the electric dipole moment of diatomic molecules using only the atomic properties of the constituent atoms. Our model can screen the entire periodic table and identify the molecules with the largest dipole moment for applications in cold molecular sciences, or find the molecule with the largest dipole moment that contains a given atom. Similarly, our model identifies useful trends that explain the dipole moment of molecules, improving our intuition in chemical physics. Finally, we condense our model into an analytical expression to predict the dipole moment in terms of atomic properties.

\end{abstract}

\maketitle

%%% Intro %%%%
The electric dipole moment is an intrinsic property or fingerprint of molecules; hence, every molecule has a different one. In cold molecular sciences, the search for diatomic molecules with large dipole moments is one of the most active research areas due to its prospective applications in dipolar quantum gases, the development of quantum gates based on polar molecules~\cite{computationdipolar,pendular2013,DeMille2002,Bohn2017,reviewquantumsimulations,Perez-Rios2020}, implementation of new Hamiltonians and the study of many-body physics~\cite{quantum1,Micheli2006,JPR2010,Carr2009,Jaksch2018,Robust,Smucker2024}, among others. Similarly, the search for new physics with molecules can benefit from molecules with larger dipole moments, especially those containing radioactive atoms like Fr-Y for hadronic charge-parity violation searches~\cite{LasercoolingRadioactivemolecules,hadronic} and Ra-Y for measuring the electric dipole moment of the electron as a consequence of charge-parity-time-reveral symmetry violation ~\cite{RaF,RaF2}. 

The dipole moment of molecules is either experimentally determined via spectroscopic techniques or calculated via ab initio quantum chemistry techniques. However, neither of these routes is a plausible approach for determining the dipole moment of the possible 6903 polar diatomic molecules. In principle, one can resort to the chemical intuition to find molecules with the largest difference in the electronegativity within the atoms and, with it, the largest dipole moments due to a larger ionic character~\cite{Pauling_paper,PaulingBook,Hannay1946,Klessinger1970}. However, it has recently been found that X-Ag molecules, where X is an alkali-metal atom, exhibit huge dipole moments, which is unexpected base on the predicted minor ionic character these molecules should show since the two atoms have similar electronegativities~\cite{LasercoolingRadioactivemolecules,Tomza_Ag}. On the basis of these results, it certainty is not a trivial task to find out which diatomic molecules can show a large dipole moment, even though this is supposed to be one of the simplest exercises in chemical physics. 

In this Letter, we employ a machine learning approach to predict the dipole moment of diatomic molecules based on atomic properties. As a result, it is possible to screen the whole set of polar diatomic molecules and find which molecules have the largest dipole moment, or to find the molecule containing a given atom type that shows the largest dipole moment. In the same vein, our machine learning models help us identify unforeseen patterns in the dipole moments of molecules based on groups and periods, which are tested against high-level ab initio electronic structure methods and confirmed. Therefore, a machine can teach us more about the nature of the dipole moment and its interconnection with essential atomic properties than anticipated. In line with these findings, we use symbolic regression to identify a simple, data-driven relationship that predicts the dipole moment with an accuracy of better than one Debye.

%%% Body of the paper %%%%

As introduced by Pauling~\cite{Pauling_paper}, the polarity of a bond is a direct consequence of atoms having different specificities to attract electrons, known as electronegativity $\chi$. Therefore, it is possible to describe the partial ionic character in terms of the electronegativity, and surprisingly enough there is no unique way to do so. Pauling proposed 
\begin{equation}
\label{eq:Pauling}
 \text{IC}_P = 100\left(1-e^{-\frac{|\chi_1-\chi_2|^2}{4}} \right), 
\end{equation}
where $\chi_i$ is the electronegativity of the $i$-th atom in the molecule. Alternatively, Hannay and Smyth~\cite{Hannay1946} proposed 
\begin{equation}
\label{eq:IC_Hanny}
 \text{IC}_{HS} = 16|\chi_1 - \chi_2| + 3.5 |\chi_1 - \chi_2|^2. 
\end{equation}
Both of these definitions, Eqs.~(\ref{eq:Pauling}) and (\ref{eq:IC_Hanny}) lead to very different partial ionic characters, even thought that both of them were empirically derived~\footnote{The main difference is in HF since Pauling assumed a 63~$\%$ of partial ionic character whereas Hanny and Smyth assumed a 41~$\%$.}.

% as suggested by Eq.~(\ref{eq1})

\begin{figure}
    \centering
    \includegraphics[width=0.9\linewidth]{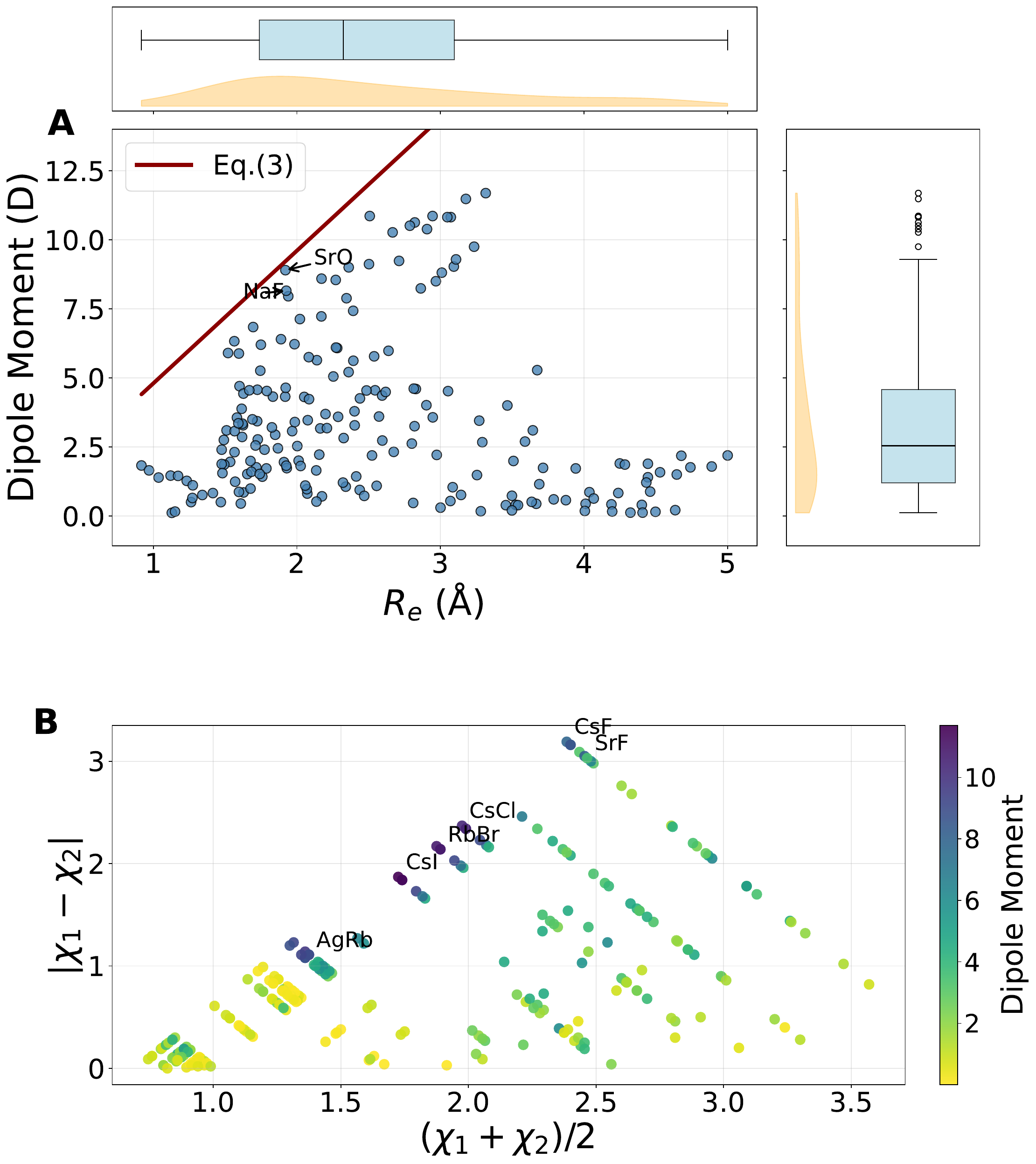}
    \caption{The dipole moment of diatomic molecules dataset containing experimental and accurate theoretical data. Panel A shows the dipole moment (in Debye)  versus the equilibrium distance (in Angstrom), as well as the maximal predicted values from Eq.~\eqref{eq1} (for $q=e$) as a dotted line and the 2 molecules with values closest to this prediction. Additionally it shows box plots for the dipole moment and equilibrium distance. The box plot shows the minimum, the maximum, the sample median, and the first and third quartiles of a given distribution. Panel B displays the Arkel–Ketelaar's triangle of the dataset, as characteristic in inorganic chemistry to characterize metallic (left corner), ionic (top corner), or covalent compounds (right corner).}
    \label{fig1}
\end{figure}

Additionally, as shown beautifully in the seminal work of Coulson~\cite{CoulsonBook}, the dipole moment has contributions beyond the bond polarity, such as the homopolar dipole moment, in relation to the size of the atom, or the anisotropy of the atomic orbitals. Therefore, the relationship between the dipole moment and the partial ionic character may not be as straightforward as thought, and certainly may depend on properties beyond the electronegativity. However, the idea that electronegativity difference of atoms within a molecule yields a large ionic character, and with it a large dipole moment, is that spread in modern chemistry that even advanced large language models fail to predict the dipole moment of molecules. For instance ChatGPT will give us ``Francium fluoride (FrF) is predicted to have the largest dipole moment, due to the extreme electronegativity difference (F being the highest, Fr the lowest) and the longer bond length from francium’s large atomic radius.", when we ask about the diatomic molecule with the largest dipole moment. Hence, a more complete understanding on the nature of the dipole moment is required and its relationship with the essential atomic properties of the atoms. 

The dipole moment, as it is thought in elementary chemistry courses, is defined as
\begin{equation}
\label{eq1}
d=qR_e,
\end{equation}
\noindent
where $q$ is the effective separated charge and $R_e$ denotes the equilibrium bond length of the molecule. As introduced by Pauling, the dipole moment is related to the polarity of molecular bonds, describing the partial ionic character of the bond, given by 
\begin{equation}
\label{eq2}
\text{IC}_d=100\frac{d}{eR_e},
\end{equation}
where $e$ is the electron charge, and the result is given in $\%$. 

%First, we noticed that the expected behavior of the dipole moment with respect the equilibrium distance, as given by Eq.~(\ref{eq1}) is not fulfilled, as shown in panel A of Fig.~\ref{fig1}. Therefore, the relationship between the dipole moment and the equilibrium distance is not linear for some molecules.

Here, we pursue a machine learning approach. To learn the dipole moment of diatomic molecules we compile a dataset comprising of 273 molecules. Of these, 140 have experimentally measured dipole moments~\cite{hwwm-1mn7,Zhang2015AuCl,10.1063/1.3226672,Zhuang2014CoO,Koelemay2023FeC,Steimle2006FeH,DeLeeuw1973CoF,10.1063/1.4794049,10.1063/1.3505141,10.1063/1.2778427,10.1063/1.465503,10.1063/1.4734596,10.1063/1.469750,10.1063/1.1822917,10.1063/1.2711807,10.1063/1.2145880,10.1063/1.3595469}, while the remaining 133 have theoretically calculated dipole moments~\cite{Tomza_Ag,Tomza-2024,Tomza-Cr,Tomza_2021} using the gold standard in quantum chemistry: Coupled clusters with singles, doubles and perturbatively triple excitations CCSD(T), which is adequate for the determination of most of the diatomics~\cite{LiuCCSDT}. First, we noticed that the relationship between the dipole moment and the equilibrium distances, using Eq.(\ref{eq1}) yields a rather disperse value of $q$, as shown in panel A of Fig.~\ref{fig1} and certainly it deviates from the extreme case of $q=e$, except NaF and SrO.  Therefore, the relationship between the dipole moment and the equilibrium distance is not linear at all for some molecules. To explore the variety of bonding and chemical behavior within this set, we employ the Van Arkel-Ketelaar Triangle~\cite{ALLEN1993647,Ketelaar1953ChemicalConstitution}, as customary in inorganic chemistry to identify the nature of chemical compounds, as shown in panel B of Fig.~\ref{fig1}. Molecules closer to the tip of the triangle are those with a larger ionic character (ionic bond), the ones in the right corner show a covalent bond and the ones in the left corner are van der Waals molecules~\footnote{Note that in the original Van Arkel-Ketelaar Triangle, compounds in the left corner are classified as metallic rather than van der Waals.}  The data set shows a relatively uniform distribution across these three bonding types, with a slight skew toward van der Waals-type bonding. Interestingly, we noticed that some of the molecules far from the tip of the triangle show a large dipole moment even though its expected ionic character is not large, contradicting the general chemical intuition and classification in chemistry. In particular, the molecules CsI (11.69 Debye), RbBr (10.86 Debye), and CsCl (10.387 Debye) have much larger dipole moments compared to the ionic bond molecules such as CsF (7.88 Debye), SrF (3.47 Debye). What is more remarkable is that there are molecules with a low difference in electronegativity that have comparable dipole moments to ionic bond molecules, such as AgRb (9.03 Debye) and AgCl (5.62 Debye). 

\begin{figure}[h]
    \centering
    \includegraphics[width=1\linewidth]{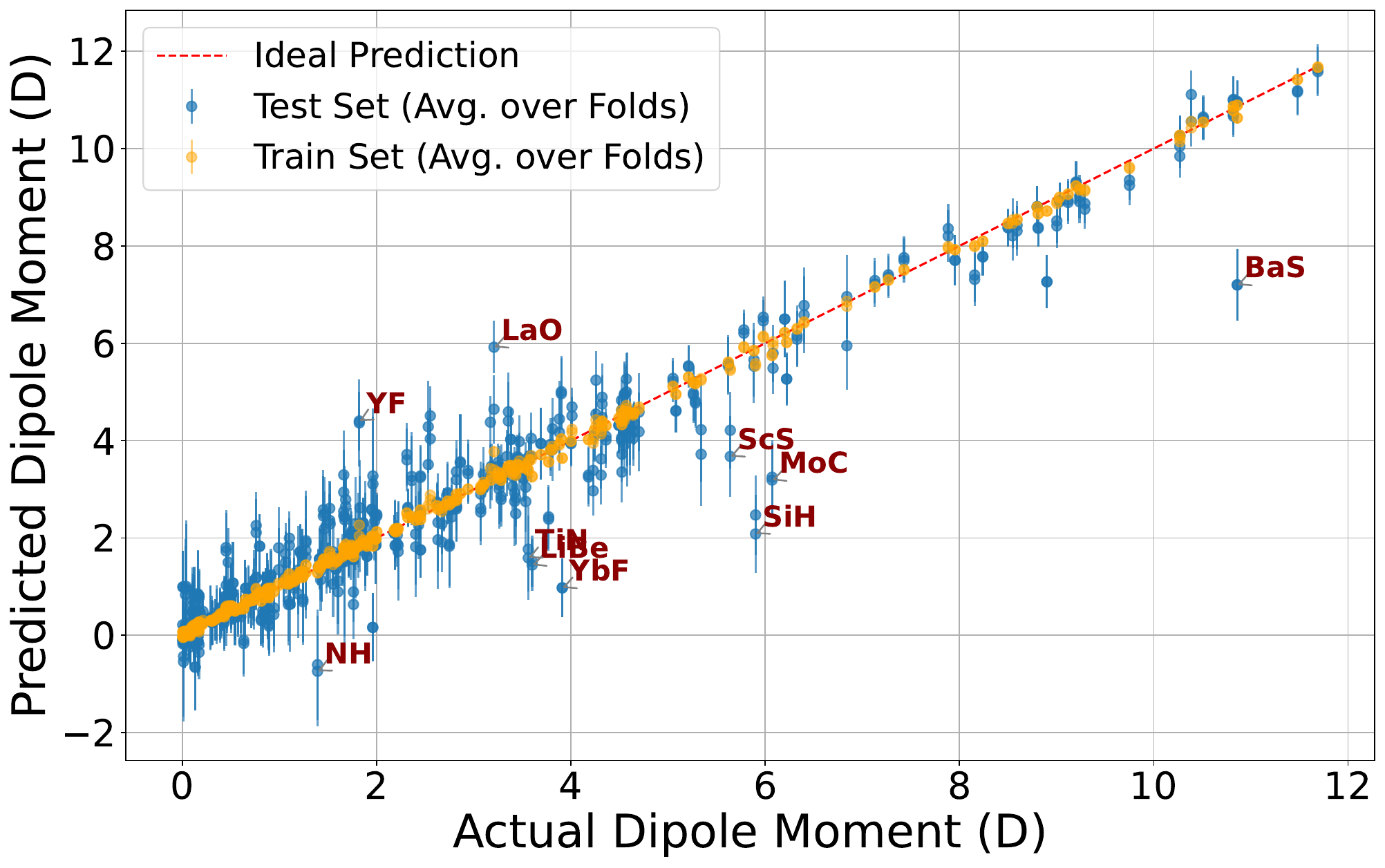}
    \caption{
        Prediction of the dipole moment of diatomics. Predicted dipole moment versus the real dipole moment. 
        The error bars represent the associated uncertainty for each prediction, and the dashed red line represents the perfect prediction. 
        The color scheme indicates whether a point belongs to the training set (orange) or the testing set (blue). 
        The labeled points are outliers with $|\mathrm{Predicted} - \mathrm{Actual}| > 1.5$ Debye.
    }
    \label{fig:dipole-prediction}
\end{figure}

We employed Gaussian Process Regression (GPR) as one of our primary predictive model due to its inherent advantages in low-data settings~\cite{williams2006gaussian}. GPR provides not only point predictions but also uncertainty estimates, which are valuable when applying the model to extrapolate. Additionally, we compare our GPR models against baseline models including Support Vector Regression (SVR), random forest, and Gradient Boosting Regression (GBR), using a consistent validation scheme to ensure a fair comparison. To validate our models, we adopted a Monte Carlo Cross-Validation (MCCV) scheme~\cite{liu2020universality}, modified to prevent data leakage from duplicated entries. In each iteration, we randomly split the dataset into 90\% training and 10\% testing. This split was done based on the Molecule's name (both the forward and backward versions have the same name in our dataset), ensuring that both the original and swapped entries for any given molecule were confined to the same split. This approach prevents information about a molecule in the training set from benefiting predictions in the test set. For all these models, we utilize a modified ionic character provided by Eq.~(\ref{eq:IC_Hanny}) $\sqrt{IC_{HS}}$  and the difference in ionization potentials (IP) $\Delta_{IP}^{1/4}$  where $\Delta_{IP}=|IP_A-IP_B|$ and also the reciprocal product of the ionization potentials $\frac{1}{IP_AIP_B}$ and the sum of the  electron affinities (EA) $(EA_A+EA_B)$ for both atoms as numerical features. Additionally, we include the group and period of the atoms within the molecule as categorical features.

\begin{table}[h]
    \centering
    \renewcommand{\arraystretch}{1.5}
    \begin{tabular}{lcc}
        \hline
        \textbf{ML method} & \textbf{RMSE} & \textbf{MAE} \\
        \hline
        \hline
        Gaussian Process Regression & 0.67 & 0.45 \\
        Gradient Boosting & 1.00 & 0.70 \\
        Random Forest & 1.05 & 0.70 \\
        Support Vector Regression & 1.07 & 0.77 \\
        \hline
    \end{tabular}
    \caption{Comparison of regression methods based on RMSE and MAE (in Debye).}
    \label{tab:method_comparison}
\end{table}

We ran 1000 iterations within the MCCV procedure and computed the average Root Mean Square Error (RMSE) and Mean Absolute Error (MAE) across all runs. The model's performance, as measured by the MAE and RMSE, is presented in Table~\ref{tab:method_comparison}.  Therefore, the model only depends on the atomic properties of the atoms. GPR outperforms the other models, with an RMSE of 0.67 Debye, followed by Gradient Boosting Regression. Therefore, from now on, we will focus on the GPR model, and its results are shown in Fig.~\ref{fig:dipole-prediction}. The model presents a few outliers like BaS, MoC, and some van der Waals molecules like LiNa, NaCs, and LiCs, also outliers even when molecular properties are included; possibly related to an anomalous natural bond orbital atomic charge versus the Mulliken charge for each of these molecules~\cite{Liudipole}.

\begin{figure}
    \centering
    \includegraphics[width=1\linewidth]{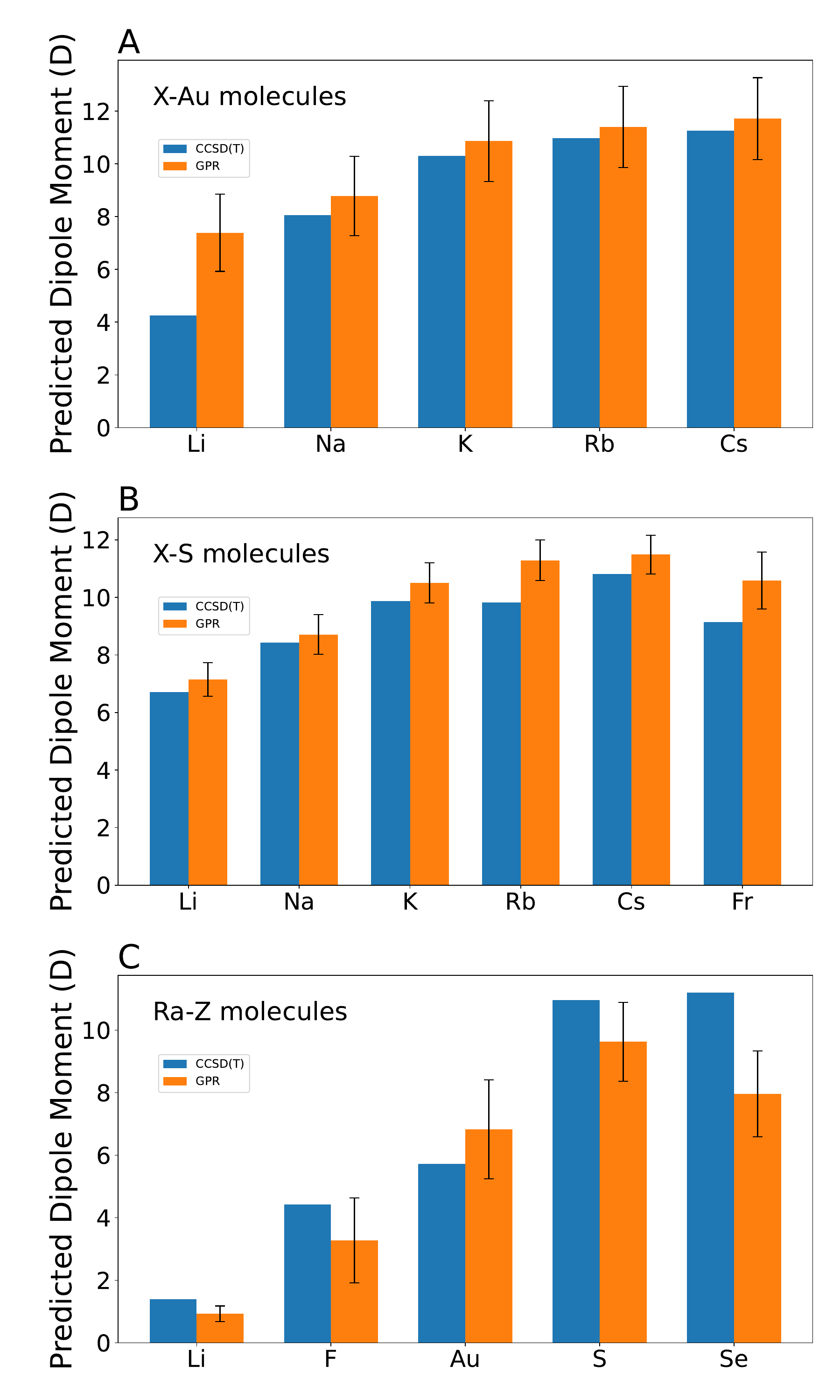}
    \caption{Testing the machine learning model predictions on 4851 molecules. Panels A and B show the predicted dipole moments for X-Au and X-S molecules (X = Alkali-metal) against CCSD(T), respectively. The machine learning model has never seen these molecules. Lastly Panel C shows Theoretical Dipole Moment for radium-containing Molecules (Ordered by CCSD(T) Smallest to Largest) versus GPR predictions. Across all 3 the trends of CCSD(T) are captured in the GPR predictions.}
    \label{Fig3}
\end{figure}

Fueled by the accuracy of the model, we search for the diatomic with the largest dipole moment. Since our model requires the ionic character and, with it, the electronegativites of the atoms, we screen 4851 molecules, as only 99 elements have reliable electronegativities within the Pauling scale. The results of our model are shown in Fig.~\ref{Fig3}, in which some unseen molecules by the model are contrasted against CCSD(T) calculations. First, we notice that the model is surprisingly accurate in predicting the dipole moment of unseen molecules in comparison with CCSD(T) calculations, even when some molecules are in the extrapolation regime, i.e., molecules unseen by the model, where machine learning techniques are supposed to lose accuracy. This confirms the robustness of our machine learning model. More importantly, the machine correctly captures the trend for X-Au and X-S molecules, where X is an alkali atom. In other words, the machine is capable of revealing the underlying correlations between the dipole moment and the atomic properties of the atoms forming the molecule.

Recently, it has been demonstrated that alkali-Ag molecules exhibit a significant dipole moment, despite the small electronegativity difference between Ag and alkali atoms. This result is very puzzling. However, as we show in panel A of Fig.~\ref{Fig3}, alkali-Au molecules show the same trend. Indeed, the dipole moment of alkali-Au molecules is larger than that of alkali-Ag molecules--another example of molecules with a relatively small electronegativity difference showing a large dipole moment. Why is that? The reason is that the group 11 atoms are transition metals with a hole on the valence shell, so they can be seen as the {\it halogens} of the transition metals. As any halogen, the largest dipole is when those are in contact with alkali atoms, hence explaining why Cu-, Ag-, and alkali-Au have a large dipole moment. Furthermore, alkali-Au molecules show a larger dipole moment than alkali-Ag molecules because Au is larger than Ag, inducing a larger homopolar dipole moment. Following that trend, alkali-Rg may show even larger dipole moments from any of the group 11 elements combined with alkali atoms, although relativistic contributions may affect the dipole moment and ending up comparable to I-alkali molecules. On the other hand, in panel B of Fig.~\ref{Fig3}, we notice that alkali-S molecules show a pretty significant dipole moment, which should not come a surprise since the alkali-O molecules show a decent dipole moment, keeping in mind that S shows a larger homopolar dipole moment than O. Similarly, if one visualizes S as a p-shell atom with two holes, it should be preferable for them to find a partner with two valence electrons, and those are alkaline-earth atoms. Indeed, S-alkaline-earth molecules show a very large dipole moment.

\setlength{\tabcolsep}{6pt}
\begin{table}[h]
  \centering
  \small
  \renewcommand{\arraystretch}{1.3}
  \begin{tabular}{l c c c}
    \toprule
    Molecule & CCSD(T) & GPR Prediction & Experimental \\
    \midrule
    CsI   & 11.73  & $11.52 \pm 0.49$  & 11.69     \\
    CsAt  & 11.72  & $10.42 \pm 0.94$ & N/A       \\
    FrAt  & 11.64  & $9.71 \pm 1.12$  & N/A       \\
    FrI   & 11.63  & $10.92 \pm 0.85$ & N/A       \\
    RbI   & 11.41  & $11.15 \pm 0.46$  & 11.48     \\
    CsAu  & 11.25  & $11.82 \pm 0.67$ & N/A       \\
    RbAu  & 10.97  & $11.76 \pm 0.86$ & N/A       \\
    CsS   & 10.82  & $11.49 \pm 0.93$ & N/A       \\
    KAu   & 10.30  & $11.1 \pm 0.85$ & N/A       \\
    FrAu  & 9.93   & $11.53 \pm 0.98$ & N/A       \\
    \bottomrule
  \end{tabular}
  \caption{Top 10 molecules with high predicted dipole moments (GPR) and available reference values.}
  \label{tab:TopDipolePredictions}
\end{table}

Another possibility that offer our model is to find the molecule with the largest dipole moment containing a given atom. To test this capability, we predict the radium-containing molecules with the largest dipole moment, and the results against CCSD(T) calculations are shown in panel C of Fig.~\ref{Fig3}. Again, the GPR model is unexpectedly well in predicting the dipole moment, although RaSe seems a little bit hard for the model. More importantly is that thanks to this exploration, we can confirm that RaS or RaSe appear as good candidates for the study of radium-containing molecules in the cold regime, and their laser cooling possibilities will be the object of future work.

After the exhaustive screening, we identify the top 10 molecules with the largest dipole moment, and the results are shown in Table~\ref{tab:TopDipolePredictions} in comparison with CCSD(T) calculations and experimental data, when available. The GPR results are pretty good in comparison with CCSD(T) calculations. The molecule with the largest dipole moment will be the combination between the heavy halogens with alkali atoms X-I or X-At, followed by alkali-Au molecules, which can be seen as the combination of the heaviest halogen transition metal atom and alkali.

To complement our machine learning models, we also applied symbolic regression via PySR \cite{cranmerInterpretableMachineLearning2023} to find an analytical formula for the dipole moment of a molecule, AB, in terms of atomic properties
\begin{equation}
d=|\Delta_{IP}^{1/4} + \left( (EA_A+EA_B) \cdot \left( \sqrt{IC_{HS}} \cdot \left( \frac{24.43}{IP_AIP_B} - 0.14 \right) \right) \right)|.
\end{equation}
The symbolic model achieved an RMSE of 1.26 Debye on the original test data, and when used on the same set of 4851 theoretical molecules, it produced an RMSE of 0.92~Debye, capturing many of the same trends as the machine learning model with reduced precision. The symbolic model offers interpretability and analytic insight that is not as visible when using GPR alone. In other words, we found the most reliable expression for the dipole moment of diatomics.

In summary, we have demonstrated through machine learning that the general idea that the difference in electronegativity determines the polarity of the molecular bond is not the whole picture, as it can lead to erroneous interpretations of the bond in diatomic molecules and fails to predict the dipole moment of diatomics. Thanks to machine learning, we are able to uncover unexpected correlations between the dipole moment and atomic properties of the constituent atoms that help us better understand the structure of the periodic table and the nature of the dipole moment. In other words, a machine can teach us some basic chemistry. In a more practical sense, and answering the question of the title of this Letter, thanks to a fast screening of diatomic molecules via machine learning techniques, we conclude that the molecules with the largest dipole moment are heavy halogens combined with heavy alkali atoms and Au-alkali molecules, all of them showing dipole $\gtrsim$11 Debye. Furthermore, we provide an analytical expression for the dipole moment of diatomics based on symbolic regression, which is the most accurate expression to date for predicting the dipole moments of any diatomic molecule. Finally, it is worth emphasizing that our machine learning model can predict the dipole moment of molecules containing a specific atom, thereby opening a new avenue for identifying the optimal molecule for testing physics beyond the standard model.

The authors acknowledge the support of the support of the United States Air Force Office of Scientific Research [grant number FA9550-23-1-0202]. M.T. acknowledges the National Science Centre Poland (grant
no.~2021/43/B/ST4/03326) for financial support.

\newpage
\newpage

\section{Supplemental information}

\subsection*{Machine Learning Methods}

In Gaussian Process Regression (GPR), the three most important factors that determine a model's success are data procurement, feature selection, and the choice of kernel. In this section, we provide a brief overview of how and why we chose to address each of these components in the way that we did.

\textbf{Data Procurement:} Our dataset consists of diatomic molecules, represented by the atomic properties of each constituent atom (e.g., electron affinity, mass, polarizability). To eliminate directional bias—i.e., the model learning to favor “atom 1” over “atom 2”—we applied a symmetrization strategy by duplicating each molecule with the atomic indices swapped. This ensures that each molecular pair appears twice in the dataset in reversed order, removing dependence on atom labeling and promoting better generalization.

\textbf{Feature Selection:} Feature selection was performed through a combination of physical intuition and statistical techniques. From the outset, we aimed for a model that could predict dipole moments of diatomic molecules using only atomic properties (hence avoiding any molecular properties and making the process of extrapolation cleaner). Initially, we included a slew of different atomic properties (electron affinity, ionization potential, electronegativity, polarizability, atomic masses, atomic radii, etc.). However, this approach proved both computationally expensive and less accurate, as it burdened the model with irrelevant information.

To filter for meaningful features, we first performed Principal Component Analysis (PCA) and observed that removing certain features such as polarizability and atomic radii—increased model performance (i.e., decreased RMSE). We then introduced categorical features (periods and groups), explicitly marking them as categorical rather than numerical to aid GPR in learning. This adjustment further improved model performance. Lastly, since we wanted symmetry to be a priority, we created symmetric modifications of important features (e.g., sum, product, absolute difference), and with the help of symbolic regression, we arrived at the final feature set. These steps resulted in a model that is not only more robust and faster but also capable of extrapolating and identifying qualitative physical trends.

\textbf{Kernel Choice:} With the explicit goal of extrapolation in mind, our kernel choice may initially appear counterintuitive: the Radial Basis Function (RBF) kernel. The RBF kernel is defined as:

\[
k_{\text{RBF}}(x, x') = a^2 \exp\left(-\frac{\|x - x'\|^2}{2\ell^2}\right),
\]

where:
\begin{itemize}
    \item $x$ and $x'$ are input data points,
    \item $a^2$ is the amplitude parameter,
    \item $\ell$ is the lengthscale hyperparameter,
    \item $\exp$ is the exponential function, and
    \item $\|x - x'\|^2$ is the squared Euclidean distance between $x$ and $x'$.
\end{itemize}

While this kernel is typically used for interpolation problems, excelling at capturing local trends but struggling with extrapolation, it performs well in our case. In particular, the total space of all possible diatomic molecules of interest is limited to

\[
\binom{118}{2} = 6903,
\]

so problems that may plague RBF kernels (e.g., short lengthscales or poor generalization) are less detrimental here. Additionally, while we have a small dataset we made sure that it was chemically diverse (as seen in the Van-Arkel-Ketelaar triangle) ensuring that the model is aware of different bond types and their assiocated dipole moments. 

We compared the RBF kernel to others designed for better extrapolation (e.g., linear, polynomial, and dot-product kernels) and found that they yielded qualitatively similar results and led to the same physical conclusions. However, the RBF kernel consistently produced the most numerically accurate predictions when benchmarked against high-level quantum chemistry results [e.g., CCSD(T)]. The only modification we made was to include a constant kernel alongside the RBF kernel for improved scaling. In conclusion, in spite of its usual shortcomings for extrapolation RBF remains the best kernel choice for our particular problem because of its relatively small size.

\begin{figure}[h]
    \centering
    \includegraphics[width=1\linewidth]{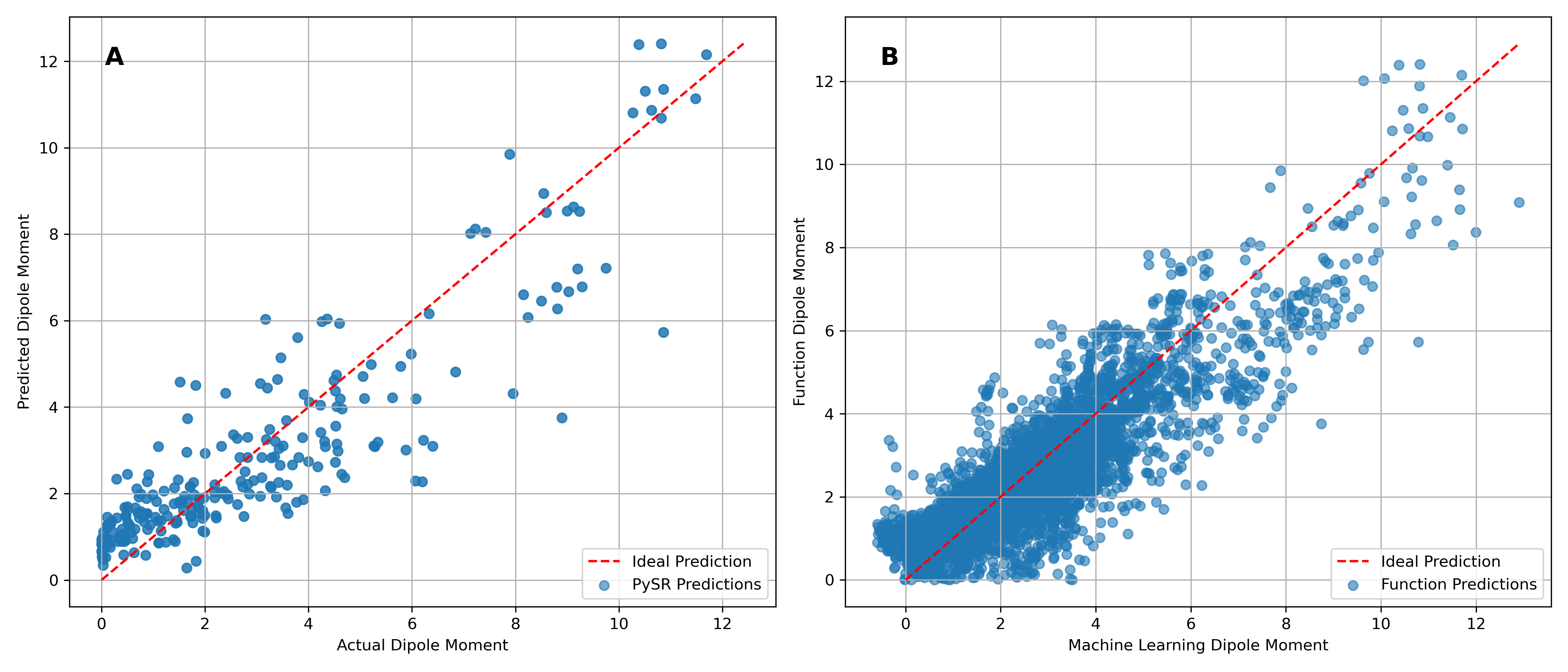}
    \caption{Figure A displays the PySR analytical model predictions on the y-axis versus the true dipole moment values from the original dataset. Figure B displays the PySR analytical model’s predictions on the y-axis versus the predicted values from the machine learning set of 4,851 molecules.}
    \label{fig:pysr_vs_datasets}
\end{figure}

\begin{table*}[ht]
    \centering
    \renewcommand{\arraystretch}{1.5}
    \begin{tabular}{|p{9cm}|c|c|}
        \hline
        \textbf{Feature Set} & \textbf{RMSE} & \textbf{MAE} \\
        \hline
        Base & 0.8334 & 0.5903 \\
        Base + Polarizability1/2 + Mass1/2 + Radius1/2 & 0.6984 & 0.4868 \\
        Base + Polarizability1/2 + Reduced Mass + Radius1/2 & 0.7075 & 0.4956 \\
        Base + Reduced Mass & 0.8750 & 0.6129 \\
        Base - Group & 1.1730 & 0.9418 \\
        Base - Period - Group & 0.8807 & 0.5918 \\
        Base - Period - Group - Ionic Potential1/2 & 1.2546 & 0.8431 \\
        Base - Period - Group - Electron Affinity1/2 & 1.1820 & 0.8250 \\
        Electronegativity1/2 + Ionic Character & 1.9249 & 1.4057 \\
        Ionic Character & 2.5642 & 1.7796 \\
        Electronegativity1/2 & 1.9066 & 1.4218 \\
        Final & 0.6670 & 0.4500 \\
        \hline
    \end{tabular}
    \caption{Comparison of feature sets with RMSE and MAE performance metrics using GPR with the same kernel. \textbf{Base} includes: Electronegativities ($EA_A$, $EA_B$), Ionization Potentials ($IP_A$, $IP_B$), Hannay and Smyth Ionic Character ($IC_{HS}$), Periods ($p_A$, $p_B$), and Groups ($g_A$, $g_B$). Final is the final model used in the paper.}
    \label{tab:feature_comparison}
\end{table*}

\subsection*{Quantum chemistry calculations}

\subsubsection{CCSD(T) calculations}
The equilibrium bond lengths and dipole moments were calculated at the CCSD(T) level of theory (coupled-cluster with singles, doubles, and perturbative triples) as implemented in the MOLPRO package \cite{molpro}. For the calculation of the dipole moment, the finite field method was employed. 

To ensure a consistent treatment of electron correlation, we employed Dunning’s augmented correlation-consistent basis sets (aug-cc-pVXZ). When core-valence correlation was necessary, we used the polarized weighted core-valence variants (aug-cc-pwCVXZ). For each atom in the system, the appropriate zeta quality was determined through systematic testing to achieve accurate property descriptions.

Additionally, for heavy elements, we used pseudopotentials to describe the core electrons, incorporating scalar relativistic effects accurately. Specifically, for atoms heavier than potassium (K), we applied the Stuttgart energy-consistent effective core potentials (ECPs) in combination with the aug-cc-pVXZ-PP basis sets optimized for these pseudopotentials. For gold, a set of additional $f$ orbitals with exponents (1.41,0.47,0.15) and $g$ orbitals with exponents (1.2,0.4) were adding along with a contraction re-scaling of 1.5.

The accuracy of the basis sets was assessed by calculating the polarizability and first ionization potential (IP) of the atomic species. The results, along with the specific basis sets used for each atom and the corresponding reference values, are presented in Table~\ref{tab:CCSDT1} . Additionally, we computed the equilibrium bond lengths and dipole moments of selected molecules previously reported in the literature, in order to benchmark the performance of the chosen methodology against prior theoretical studies and available experimental data. The results of this molecular benchmarking are summarized in Table~III. 

\begin{table*}
    \centering
    \renewcommand{\arraystretch}{1.5}
    \begin{tabular}{cc|cc|cc|cc|cc|cc|cc|}
        \multicolumn{2}{c|}{    \textbf{    Atom }      } & \multicolumn{2}{c|}{    \textbf{    ECP }      } & \multicolumn{2}{c|}{    \textbf{    Basis set }      } & \multicolumn{4}{c|}{    \textbf{    Polarization (au)}      } 
                     & \multicolumn{4}{c|}{    \textbf{    IP (cm$^{-1}$) }      }
        \\
                \multicolumn{2}{c|}{    \textbf{    }      } & \multicolumn{2}{c|}{    \textbf{    }      } & \multicolumn{2}{c|}{    \textbf{     }      } &  \multicolumn{2}{c}{    \textbf{  REF. \cite{Polarizabilities}  }      } & \multicolumn{2}{c|}{    \textbf{  Calculated  }      } 
                     &  \multicolumn{2}{c}{    \textbf{ REF. \cite{NIST_ASD}  }      } & \multicolumn{2}{c|}{    \textbf{  Calculated  }      }
        \\
        \hline
       \multicolumn{2}{c}  \textbf{Li} & \multicolumn{2}{c} {-}& \multicolumn{2}{c} {aug-cc-pwCVQZ
}& \multicolumn{2}{c} {164.1125}& \multicolumn{2}{c} {166.106} & \multicolumn{2}{c} {43488.117}& \multicolumn{2}{c} {43120.332}\\
       \multicolumn{2}{c}  \textbf{Na} & \multicolumn{2}{c} {-}& \multicolumn{2}{c} {aug-cc-pwCV5Z
}& \multicolumn{2}{c} {162.7}& \multicolumn{2}{c} {163.569} & \multicolumn{2}{c} {41449.430}& \multicolumn{2}{c} {41391.594}\\
       \multicolumn{2}{c}  \textbf{K} & \multicolumn{2}{c} {-}& \multicolumn{2}{c} {aug-cc-pwCVQZ
}& \multicolumn{2}{c} {294.3}& \multicolumn{2}{c} {297.8} & \multicolumn{2}{c} {35010.621}& \multicolumn{2}{c} {34930.985}\\
       \multicolumn{2}{c}  \textbf{Rb} & \multicolumn{2}{c} {ECP28MDF}& \multicolumn{2}{c} {aug-cc-pwCV5Z-PP
}& \multicolumn{2}{c} {319.8}& \multicolumn{2}{c} {319.6} & \multicolumn{2}{c} {33691.587}& \multicolumn{2}{c} {33622.313}\\
       \multicolumn{2}{c}  \textbf{Cs} & \multicolumn{2}{c} {ECP46MDF}& \multicolumn{2}{c} {aug-cc-pwCV5Z-PP
}& \multicolumn{2}{c} {400.9}& \multicolumn{2}{c} {397.622} & \multicolumn{2}{c} {31407.192}& \multicolumn{2}{c} {31389.254}\\
       \multicolumn{2}{c}  \textbf{Fr} & \multicolumn{2}{c} {ECP78MDF}& \multicolumn{2}{c} {aug-cc-pwCV5Z-PP
}& \multicolumn{2}{c} {317.8}& \multicolumn{2}{c} {325.70} & \multicolumn{2}{c} {37932.239}& \multicolumn{2}{c} {32431.540}\\
       \multicolumn{2}{c}  \textbf{Ca} & \multicolumn{2}{c} {ECP10MDF}& \multicolumn{2}{c} {aug-cc-pwCV5Z-PP
}& \multicolumn{2}{c} {160.8}& \multicolumn{2}{c} {159.28} & \multicolumn{2}{c} {49307.057}& \multicolumn{2}{c} {49362.752}\\
       \multicolumn{2}{c}  \textbf{Sr} & \multicolumn{2}{c} {ECP28MDF}& \multicolumn{2}{c} {aug-cc-pwCV5Z-PP
}& \multicolumn{2}{c} {197.2}& \multicolumn{2}{c} {198.907} & \multicolumn{2}{c} {45933.284}& \multicolumn{2}{c} {45854.457}\\
       \multicolumn{2}{c}  \textbf{Ba} & \multicolumn{2}{c} {ECP46MDF}& \multicolumn{2}{c} {aug-cc-pwCV5Z-PP
}& \multicolumn{2}{c} {272.0}& \multicolumn{2}{c} {274.57} & \multicolumn{2}{c} {42035.842}& \multicolumn{2}{c} {41895.015}\\
       \multicolumn{2}{c}  \textbf{Ra} & \multicolumn{2}{c} {ECP78MDF}& \multicolumn{2}{c} {aug-cc-pwCV5Z-PP
}& \multicolumn{2}{c} {246}& \multicolumn{2}{c} {254.2} & \multicolumn{2}{c} {42574.342}& \multicolumn{2}{c} {41507.561}\\
       \multicolumn{2}{c}  \textbf{S} & \multicolumn{2}{c} {-}& \multicolumn{2}{c} {aug-cc-pwCV5Z
}& \multicolumn{2}{c} {19.4}& \multicolumn{2}{c} {19.99} & \multicolumn{2}{c} {83561.097}& \multicolumn{2}{c} {85876.445}\\
       \multicolumn{2}{c}  \textbf{Se} & \multicolumn{2}{c} {ECP10MDF}& \multicolumn{2}{c} {aug-cc-pwCV5Z-PP
}& \multicolumn{2}{c} {28.9}& \multicolumn{2}{c} {24.0d} & \multicolumn{2}{c} {78659.967}& \multicolumn{2}{c} {75737.205}\\
       \multicolumn{2}{c}  \textbf{F} & \multicolumn{2}{c} {-}& \multicolumn{2}{c} {aug-cc-pwCV5Z
}& \multicolumn{2}{c} {3.74}& \multicolumn{2}{c} {3.40} & \multicolumn{2}{c} {140527.762}& \multicolumn{2}{c} {140369.762}\\
       \multicolumn{2}{c}  \textbf{I} & \multicolumn{2}{c} {ECP46MDF}& \multicolumn{2}{c} {aug-cc-pwCVQZ-PP
}& \multicolumn{2}{c} {32.9}& \multicolumn{2}{c} {29.80} & \multicolumn{2}{c} {84292.969}& \multicolumn{2}{c} {92197.199}\\
       \multicolumn{2}{c}  \textbf{At} & \multicolumn{2}{c} {ECP78MDF}& \multicolumn{2}{c} {aug-cc-pwCVQZ-PP} & \multicolumn{2}{c} {42} &\multicolumn{2}{c} {42.33} & \multicolumn{2}{c} {75154.711}& \multicolumn{2}{c} {83075.453}\\
       \multicolumn{2}{c}  \textbf{Au} & \multicolumn{2}{c} {ECP60MDF}& \multicolumn{2}{c} {aug-cc-pwCV5Z-PP
}& \multicolumn{2}{c} {36}& \multicolumn{2}{c} {37.21} & \multicolumn{2}{c} {74410.827}& \multicolumn{2}{c} {72991.226}\\
       \multicolumn{2}{c}  \textbf{Pd} & \multicolumn{2}{c} {ECP28MDF}& \multicolumn{2}{c} {aug-cc-pwCV5Z-PP
}& \multicolumn{2}{c} {26.14}& \multicolumn{2}{c} {26.19} & \multicolumn{2}{c} {67243.991}& \multicolumn{2}{c} {67308.517}\\

 \hline
    \end{tabular}
    \caption{Basis sets details along with the values of Polarization and IP for the explored atoms in the CCSD(T) calculations. All the reference values for polarizability and IP were taken from \cite{Polarizabilities} and \cite{NIST_ASD}, respectively. }
            \label{tab:CCSDT1}
\end{table*}

\begin{table}
    \centering
    \renewcommand{\arraystretch}{1.5}
    \resizebox{\columnwidth}{!}{\begin{tabular}{cc|cc|cc|cc|cc|}
        \multicolumn{2}{c|}{    \textbf{    Molecule }      }  & \multicolumn{4}{c|}{    \textbf{    R$_{\rm{eq}}$(\AA)}      } 
                     & \multicolumn{4}{c|}{    \textbf{    Dipole (Debye) }      }
        \\
                \multicolumn{2}{c|}{    \textbf{    }      } &  \multicolumn{2}{c}{    \textbf{  REF.  }      } & \multicolumn{2}{c|}{    \textbf{  Calculated  }      } 
                     &  \multicolumn{2}{c}{    \textbf{ REF.   }      } & \multicolumn{2}{c|}{    \textbf{  Calculated  }      }
        \\
        \hline
       \multicolumn{2}{c}  \textbf{AuF}& \multicolumn{2}{c} {1.918 \cite{AuF} }& \multicolumn{2}{c} {1.917} & \multicolumn{2}{c} {4.13\cite{AuF}}& \multicolumn{2}{c} {4.33}\\
                      \multicolumn{2}{c}  \textbf{AuS}& \multicolumn{2}{c} {2.156 \cite{AuS} }& \multicolumn{2}{c} {2.187 } & \multicolumn{2}{c} {2.22 \cite{AuS}}& \multicolumn{2}{c} {2.34 }\\   
              \multicolumn{2}{c}  \textbf{SF}& \multicolumn{2}{c} {1.596 \cite{SF} }& \multicolumn{2}{c} {1.583} & \multicolumn{2}{c} {0.794 \cite{SF}}& \multicolumn{2}{c} {0.783 }\\
    \multicolumn{2}{c}  \textbf{SeF}& \multicolumn{2}{c} {1.714 \cite{SeF} }& \multicolumn{2}{c} {1.741} & \multicolumn{2}{c} {1.520
 \cite{SeF}}& \multicolumn{2}{c} {1.556 }\\         \multicolumn{2}{c}  \textbf{CsI}& \multicolumn{2}{c} {3.37 \cite{XI} }& \multicolumn{2}{c} {3.340} & \multicolumn{2}{c} {11.69
 \cite{XI}}& \multicolumn{2}{c} {11.733 }\\         \multicolumn{2}{c}  \textbf{RbI}& \multicolumn{2}{c} {3.177 \cite{XI} }& \multicolumn{2}{c} { 3.179 } & \multicolumn{2}{c} { 11.48
 \cite{XI}}& \multicolumn{2}{c} { 11.407}\\     
                   
 \hline
    \end{tabular}}
    \label{tab:CCSDT2}
    \caption{Comparison of bond lengths and dipole moments computed using our methodology in MOLPRO with reference values reported in the literature.  }
\end{table}

\subsubsection{DFT calculations}
The equilibrium bond lengths, first order vibrational harmonic frequencies, and dipole moments were calculated with density functional theory (DFT) implemented in the Gaussian 16 program~\cite{g16}. 

To achieve high accuracy, we chose the $\omega$B97X-D functional, which includes both long-range and short-range exchange correlations, and empirical dispersion corrections, and is reported to generally outperform the commonly used B3LYP functional~\cite{wb97xd}.  

For all the atoms reported in this paper, which are mostly S-block and D-block atoms, we used the recommended quadruple zeta valence basis set with two sets of polarization functions (def2-QZVPP), since for DFT dipole moment calculations, it is close to the DFT basis set limit, and for DFT geometry optimizations, it presents extremely small bases errors~\cite{def2}. Basis set data for all the atoms are downloaded from Basis Set Exchange~\cite{basis1,basis2,basis3}. The results are presented in Table~\ref{tab:CCSD_DFT_GPR_Table}.

\subsection{Machine learning predictions against quantum chemical calculations}
In Table~\ref{tab:CCSD_DFT_GPR_Table}, we present a comparison between our GPR model prediction of the dipole moment of diatomic versus CCSD(T) and density functional theory (DFT) calculations. 

\setlength{\tabcolsep}{4pt} 
\begin{table*}[h]
  \centering
  \small
  \renewcommand{\arraystretch}{1.3}
  \begin{tabular}{p{1.5cm}|c|c|c|c|c}
    \hline
    \textbf{Molecule} & \textbf{CCSD(T)} & \textbf{DFT} & \textbf{GPR} & \textbf{GPR Lower Bound} & \textbf{GPR Upper Bound} \\
    \hline
    BaAu & 4.3630 & 5.5464 & 6.02 & 5.13 & 6.92 \\
    CaAu & 3.6100 & 2.6427 & 4.55 & 3.72 & 5.38 \\
    CsAu & 11.2520 & 10.8127 & 11.82 & 10.89 & 12.74 \\
    AuF  & 4.2090 & nan    & 4.82 & 4.25 & 5.39 \\
    KAu  & 10.3050 & 9.8416 & 11.10 & 10.25 & 11.95 \\
    LiAu & 4.2450 & 6.1492 & 7.56 & 6.80 & 8.33 \\
    NaAu & 8.0510 & 7.7114 & 9.68 & 8.86 & 10.50 \\
    RaAu & 5.7260 & nan    & 6.40 & 5.46 & 7.34 \\
    RbAu & 10.9720 & 10.5441 & 11.76 & 10.90 & 12.62 \\
    AuS  & 2.3200 & nan    & 2.00 & 1.11 & 2.89 \\
    SrAu & 4.1890 & 3.6073 & 5.57 & 4.73 & 6.40 \\
    KPd  & 7.7110 & nan    & 6.33 & 5.25 & 7.41 \\
    LiPd & 4.5270 & nan    & 4.37 & 3.40 & 5.35 \\
    RbPd & 8.0290 & nan    & 6.18 & 5.09 & 7.27 \\
    RaF  & 4.4320 & nan    & 3.19 & 2.05 & 4.34 \\
    LiRa & 1.3960 & nan    & 0.427 & 0.165 & 0.756 \\
    RaSe & 11.1960 & nan   & 7.98 & 7.08 & 8.88 \\
    BaS  & 10.8850 & 11.6329 & 6.58 & 5.83 & 7.34 \\
    CaS  & 10.1480 & 11.3756 & 8.24 & 7.73 & 8.76 \\
    CsS  & 10.82 & 10.3621 & 11.49 & 10.82 & 12.20 \\
    BaSe & 11.1770 & 12.1959 & 8.59 & 7.98 & 9.20 \\
    CaSe & 10.0310 & 11.5003 & 6.71 & 6.10 & 7.35 \\
    CsSe & 11.0790 & 10.8586 & 10.58 & 9.81 & 11.34 \\
    SeF  & 1.5560 & nan    & 1.33 & 0.70 & 1.96 \\
    KSe  & 10.2350 & 10.8360 & 9.86 & 9.12 & 10.60 \\
    LiSe & 6.8400 & 7.2794 & 6.26 & 5.63 & 6.70 \\
    NaSe & 8.6190 & 8.9761 & 7.95 & 7.24 & 8.65 \\
    RbSe & 10.8310 & 11.5469 & 10.71 & 9.97 & 11.45 \\
    SrSe & 10.6540 & 12.1073 & 4.59 & 3.78 & 5.40 \\
    SF   & 0.7830 & nan    & 1.70 & 1.08 & 2.32 \\
    KS   & 9.8680 & 10.6156 & 10.51 & 9.81 & 11.21 \\
    LiS  & 6.7080 & 7.2321 & 7.15 & 6.57 & 7.73 \\
    NaS  & 8.4280 & 8.9388 & 7.95 & 7.24 & 8.65 \\
    RbS  & 9.8240 & 11.2679 & 11.29 & 10.59 & 11.99 \\
    SrS  & 10.6230 & 11.8811 & 4.76 & 3.92 & 5.60 \\
    FrS & 9.14 & nan & 10.58 & 9.60 & 11.57 \\
    \hline
  \end{tabular}
  \caption{CCSD(T) vs DFT vs GPR}
  \label{tab:CCSD_DFT_GPR_Table}
\end{table*}

%\newpage

\bibliographystyle{apsrev}
\bibliography{dipolemoment}% Produces the bibliography via BibTeX.

\begin{thebibliography}{66}
\expandafter\ifx\csname natexlab\endcsname\relax\def\natexlab#1{#1}\fi
\expandafter\ifx\csname bibnamefont\endcsname\relax
  \def\bibnamefont#1{#1}\fi
\expandafter\ifx\csname bibfnamefont\endcsname\relax
  \def\bibfnamefont#1{#1}\fi
\expandafter\ifx\csname citenamefont\endcsname\relax
  \def\citenamefont#1{#1}\fi
\expandafter\ifx\csname url\endcsname\relax
  \def\url#1{\texttt{#1}}\fi
\expandafter\ifx\csname urlprefix\endcsname\relax\def\urlprefix{URL }\fi
\providecommand{\bibinfo}[2]{#2}
\providecommand{\eprint}[2][]{\url{#2}}

\bibitem[{\citenamefont{Ni et~al.}(2018)\citenamefont{Ni, Rosenband, and Grimes}}]{computationdipolar}
\bibinfo{author}{\bibfnamefont{K.-K.} \bibnamefont{Ni}}, \bibinfo{author}{\bibfnamefont{T.}~\bibnamefont{Rosenband}}, \bibnamefont{and} \bibinfo{author}{\bibfnamefont{D.~D.} \bibnamefont{Grimes}}, \bibinfo{journal}{Chem. Sci.} \textbf{\bibinfo{volume}{9}}, \bibinfo{pages}{6830} (\bibinfo{year}{2018}), \urlprefix\url{http://dx.doi.org/10.1039/C8SC02355G}.

\bibitem[{\citenamefont{Zhu et~al.}(2013)\citenamefont{Zhu, Kais, Wei, Herschbach, and Friedrich}}]{pendular2013}
\bibinfo{author}{\bibfnamefont{J.}~\bibnamefont{Zhu}}, \bibinfo{author}{\bibfnamefont{S.}~\bibnamefont{Kais}}, \bibinfo{author}{\bibfnamefont{Q.}~\bibnamefont{Wei}}, \bibinfo{author}{\bibfnamefont{D.}~\bibnamefont{Herschbach}}, \bibnamefont{and} \bibinfo{author}{\bibfnamefont{B.}~\bibnamefont{Friedrich}}, \bibinfo{journal}{The Journal of Chemical Physics} \textbf{\bibinfo{volume}{138}}, \bibinfo{pages}{024104} (\bibinfo{year}{2013}), ISSN \bibinfo{issn}{0021-9606}, \eprint{https://pubs.aip.org/aip/jcp/article-pdf/doi/10.1063/1.4774058/14791601/024104\_1\_online.pdf}, \urlprefix\url{https://doi.org/10.1063/1.4774058}.

\bibitem[{\citenamefont{DeMille}(2002)}]{DeMille2002}
\bibinfo{author}{\bibfnamefont{D.}~\bibnamefont{DeMille}}, \bibinfo{journal}{Phys. Rev. Lett.} \textbf{\bibinfo{volume}{88}}, \bibinfo{pages}{067901} (\bibinfo{year}{2002}), \urlprefix\url{https://link.aps.org/doi/10.1103/PhysRevLett.88.067901}.

\bibitem[{\citenamefont{Bohn et~al.}(2017)\citenamefont{Bohn, Rey, and Ye}}]{Bohn2017}
\bibinfo{author}{\bibfnamefont{J.~L.} \bibnamefont{Bohn}}, \bibinfo{author}{\bibfnamefont{A.~M.} \bibnamefont{Rey}}, \bibnamefont{and} \bibinfo{author}{\bibfnamefont{J.}~\bibnamefont{Ye}}, \bibinfo{journal}{Science} \textbf{\bibinfo{volume}{357}}, \bibinfo{pages}{1002 } (\bibinfo{year}{2017}), \urlprefix\url{https://api.semanticscholar.org/CorpusID:206656605}.

\bibitem[{\citenamefont{Sch{\"a}fer et~al.}(2020)\citenamefont{Sch{\"a}fer, Fukuhara, Sugawa, Takasu, and Takahashi}}]{reviewquantumsimulations}
\bibinfo{author}{\bibfnamefont{F.}~\bibnamefont{Sch{\"a}fer}}, \bibinfo{author}{\bibfnamefont{T.}~\bibnamefont{Fukuhara}}, \bibinfo{author}{\bibfnamefont{S.}~\bibnamefont{Sugawa}}, \bibinfo{author}{\bibfnamefont{Y.}~\bibnamefont{Takasu}}, \bibnamefont{and} \bibinfo{author}{\bibfnamefont{Y.}~\bibnamefont{Takahashi}}, \bibinfo{journal}{Nature Reviews Physics} \textbf{\bibinfo{volume}{2}}, \bibinfo{pages}{411} (\bibinfo{year}{2020}), \urlprefix\url{https://doi.org/10.1038/s42254-020-0195-3}.

\bibitem[{\citenamefont{P\'{e}rez-R\'{i}os}(2020)}]{Perez-Rios2020}
\bibinfo{author}{\bibfnamefont{J.}~\bibnamefont{P\'{e}rez-R\'{i}os}}, \emph{\bibinfo{title}{An Introduction to Cold and Ultracold Chemistry}} (\bibinfo{publisher}{Springer International Publishing}, \bibinfo{address}{Cham, Switzerland}, \bibinfo{year}{2020}), ISBN \bibinfo{isbn}{3030559351}.

\bibitem[{\citenamefont{Kruckenhauser et~al.}(2020)\citenamefont{Kruckenhauser, Sieberer, De~Marco, Li, Matsuda, Tobias, Valtolina, Ye, Rey, Baranov et~al.}}]{quantum1}
\bibinfo{author}{\bibfnamefont{A.}~\bibnamefont{Kruckenhauser}}, \bibinfo{author}{\bibfnamefont{L.~M.} \bibnamefont{Sieberer}}, \bibinfo{author}{\bibfnamefont{L.}~\bibnamefont{De~Marco}}, \bibinfo{author}{\bibfnamefont{J.-R.} \bibnamefont{Li}}, \bibinfo{author}{\bibfnamefont{K.}~\bibnamefont{Matsuda}}, \bibinfo{author}{\bibfnamefont{W.~G.} \bibnamefont{Tobias}}, \bibinfo{author}{\bibfnamefont{G.}~\bibnamefont{Valtolina}}, \bibinfo{author}{\bibfnamefont{J.}~\bibnamefont{Ye}}, \bibinfo{author}{\bibfnamefont{A.~M.} \bibnamefont{Rey}}, \bibinfo{author}{\bibfnamefont{M.~A.} \bibnamefont{Baranov}}, \bibnamefont{et~al.}, \bibinfo{journal}{Phys. Rev. A} \textbf{\bibinfo{volume}{102}}, \bibinfo{pages}{023320} (\bibinfo{year}{2020}), \urlprefix\url{https://link.aps.org/doi/10.1103/PhysRevA.102.023320}.

\bibitem[{\citenamefont{Micheli et~al.}(2006)\citenamefont{Micheli, Brennen, and Zoller}}]{Micheli2006}
\bibinfo{author}{\bibfnamefont{A.}~\bibnamefont{Micheli}}, \bibinfo{author}{\bibfnamefont{G.~K.} \bibnamefont{Brennen}}, \bibnamefont{and} \bibinfo{author}{\bibfnamefont{P.}~\bibnamefont{Zoller}}, \bibinfo{journal}{Nature Physics} \textbf{\bibinfo{volume}{2}}, \bibinfo{pages}{341} (\bibinfo{year}{2006}), \urlprefix\url{https://doi.org/10.1038/nphys287}.

\bibitem[{\citenamefont{P{\'{e}}rez-R{\'{\i}}os et~al.}(2010)\citenamefont{P{\'{e}}rez-R{\'{\i}}os, Herrera, and Krems}}]{JPR2010}
\bibinfo{author}{\bibfnamefont{J.}~\bibnamefont{P{\'{e}}rez-R{\'{\i}}os}}, \bibinfo{author}{\bibfnamefont{F.}~\bibnamefont{Herrera}}, \bibnamefont{and} \bibinfo{author}{\bibfnamefont{R.~V.} \bibnamefont{Krems}}, \bibinfo{journal}{New Journal of Physics} \textbf{\bibinfo{volume}{12}}, \bibinfo{pages}{103007} (\bibinfo{year}{2010}), \urlprefix\url{https://doi.org/10.1088/1367-2630/12/10/103007}.

\bibitem[{\citenamefont{Carr et~al.}(2009)\citenamefont{Carr, DeMille, Krems, and Ye}}]{Carr2009}
\bibinfo{author}{\bibfnamefont{L.~D.} \bibnamefont{Carr}}, \bibinfo{author}{\bibfnamefont{D.}~\bibnamefont{DeMille}}, \bibinfo{author}{\bibfnamefont{R.~V.} \bibnamefont{Krems}}, \bibnamefont{and} \bibinfo{author}{\bibfnamefont{J.}~\bibnamefont{Ye}}, \bibinfo{journal}{New Journal of Physics} \textbf{\bibinfo{volume}{11}}, \bibinfo{pages}{055049} (\bibinfo{year}{2009}), \urlprefix\url{https://doi.org/10.1088/1367-2630/11/5/055049}.

\bibitem[{\citenamefont{Blackmore et~al.}(2018{\natexlab{a}})\citenamefont{Blackmore, Caldwell, Gregory, Bridge, Sawant, Aldegunde, Mur-Petit, Jaksch, Hutson, Sauer et~al.}}]{Jaksch2018}
\bibinfo{author}{\bibfnamefont{J.~A.} \bibnamefont{Blackmore}}, \bibinfo{author}{\bibfnamefont{L.}~\bibnamefont{Caldwell}}, \bibinfo{author}{\bibfnamefont{P.~D.} \bibnamefont{Gregory}}, \bibinfo{author}{\bibfnamefont{E.~M.} \bibnamefont{Bridge}}, \bibinfo{author}{\bibfnamefont{R.}~\bibnamefont{Sawant}}, \bibinfo{author}{\bibfnamefont{J.}~\bibnamefont{Aldegunde}}, \bibinfo{author}{\bibfnamefont{J.}~\bibnamefont{Mur-Petit}}, \bibinfo{author}{\bibfnamefont{D.}~\bibnamefont{Jaksch}}, \bibinfo{author}{\bibfnamefont{J.~M.} \bibnamefont{Hutson}}, \bibinfo{author}{\bibfnamefont{B.~E.} \bibnamefont{Sauer}}, \bibnamefont{et~al.}, \bibinfo{journal}{Quantum Sciences and Techonology} \textbf{\bibinfo{volume}{4}}, \bibinfo{pages}{014010} (\bibinfo{year}{2018}{\natexlab{a}}), \urlprefix\url{https://doi.org/10.1088/2058-9565/aaee35}.

\bibitem[{\citenamefont{Blackmore et~al.}(2018{\natexlab{b}})\citenamefont{Blackmore, Caldwell, Gregory, Bridge, Sawant, Aldegunde, Mur-Petit, Jaksch, Hutson, Sauer et~al.}}]{Robust}
\bibinfo{author}{\bibfnamefont{J.~A.} \bibnamefont{Blackmore}}, \bibinfo{author}{\bibfnamefont{L.}~\bibnamefont{Caldwell}}, \bibinfo{author}{\bibfnamefont{P.~D.} \bibnamefont{Gregory}}, \bibinfo{author}{\bibfnamefont{E.~M.} \bibnamefont{Bridge}}, \bibinfo{author}{\bibfnamefont{R.}~\bibnamefont{Sawant}}, \bibinfo{author}{\bibfnamefont{J.}~\bibnamefont{Aldegunde}}, \bibinfo{author}{\bibfnamefont{J.}~\bibnamefont{Mur-Petit}}, \bibinfo{author}{\bibfnamefont{D.}~\bibnamefont{Jaksch}}, \bibinfo{author}{\bibfnamefont{J.~M.} \bibnamefont{Hutson}}, \bibinfo{author}{\bibfnamefont{B.~E.} \bibnamefont{Sauer}}, \bibnamefont{et~al.}, \bibinfo{journal}{Quantum Science and Technology} \textbf{\bibinfo{volume}{4}}, \bibinfo{pages}{014010} (\bibinfo{year}{2018}{\natexlab{b}}), \urlprefix\url{https://doi.org/10.1088/2058-9565/aaee35}.

\bibitem[{\citenamefont{Smucker and P{\'e}rez-R{\'\i}os}(2024)}]{Smucker2024}
\bibinfo{author}{\bibfnamefont{J.}~\bibnamefont{Smucker}} \bibnamefont{and} \bibinfo{author}{\bibfnamefont{J.}~\bibnamefont{P{\'e}rez-R{\'\i}os}}, \bibinfo{journal}{Phys. Chem. Chem. Phys.} \textbf{\bibinfo{volume}{26}}, \bibinfo{pages}{21513} (\bibinfo{year}{2024}), \urlprefix\url{http://dx.doi.org/10.1039/D4CP01956C}.

\bibitem[{\citenamefont{Kłos et~al.}(2022)\citenamefont{Kłos, Li, Tiesinga, and Kotochigova}}]{LasercoolingRadioactivemolecules}
\bibinfo{author}{\bibfnamefont{J.}~\bibnamefont{Kłos}}, \bibinfo{author}{\bibfnamefont{H.}~\bibnamefont{Li}}, \bibinfo{author}{\bibfnamefont{E.}~\bibnamefont{Tiesinga}}, \bibnamefont{and} \bibinfo{author}{\bibfnamefont{S.}~\bibnamefont{Kotochigova}}, \bibinfo{journal}{New Journal of Physics} \textbf{\bibinfo{volume}{24}}, \bibinfo{pages}{025005} (\bibinfo{year}{2022}), \urlprefix\url{https://dx.doi.org/10.1088/1367-2630/ac50ea}.

\bibitem[{\citenamefont{Marc et~al.}(2023)\citenamefont{Marc, Hubert, and Fleig}}]{hadronic}
\bibinfo{author}{\bibfnamefont{A.}~\bibnamefont{Marc}}, \bibinfo{author}{\bibfnamefont{M.}~\bibnamefont{Hubert}}, \bibnamefont{and} \bibinfo{author}{\bibfnamefont{T.}~\bibnamefont{Fleig}}, \bibinfo{journal}{Phys. Rev. A} \textbf{\bibinfo{volume}{108}}, \bibinfo{pages}{062815} (\bibinfo{year}{2023}), \urlprefix\url{https://link.aps.org/doi/10.1103/PhysRevA.108.062815}.

\bibitem[{\citenamefont{Garcia~Ruiz et~al.}(2020)\citenamefont{Garcia~Ruiz, Berger, Billowes, Binnersley, Bissell, Breier, Brinson, Chrysalidis, Cocolios, Cooper et~al.}}]{RaF}
\bibinfo{author}{\bibfnamefont{R.~F.} \bibnamefont{Garcia~Ruiz}}, \bibinfo{author}{\bibfnamefont{R.}~\bibnamefont{Berger}}, \bibinfo{author}{\bibfnamefont{J.}~\bibnamefont{Billowes}}, \bibinfo{author}{\bibfnamefont{C.~L.} \bibnamefont{Binnersley}}, \bibinfo{author}{\bibfnamefont{M.~L.} \bibnamefont{Bissell}}, \bibinfo{author}{\bibfnamefont{A.~A.} \bibnamefont{Breier}}, \bibinfo{author}{\bibfnamefont{A.~J.} \bibnamefont{Brinson}}, \bibinfo{author}{\bibfnamefont{K.}~\bibnamefont{Chrysalidis}}, \bibinfo{author}{\bibfnamefont{T.~E.} \bibnamefont{Cocolios}}, \bibinfo{author}{\bibfnamefont{B.~S.} \bibnamefont{Cooper}}, \bibnamefont{et~al.}, \bibinfo{journal}{Nature} \textbf{\bibinfo{volume}{581}}, \bibinfo{pages}{396} (\bibinfo{year}{2020}), \urlprefix\url{https://doi.org/10.1038/s41586-020-2299-4}.

\bibitem[{\citenamefont{Udrescu et~al.}(2021)\citenamefont{Udrescu, Brinson, Ruiz, Gaul, Berger, Billowes, Binnersley, Bissell, Breier, Chrysalidis et~al.}}]{RaF2}
\bibinfo{author}{\bibfnamefont{S.~M.} \bibnamefont{Udrescu}}, \bibinfo{author}{\bibfnamefont{A.~J.} \bibnamefont{Brinson}}, \bibinfo{author}{\bibfnamefont{R.~F.~G.} \bibnamefont{Ruiz}}, \bibinfo{author}{\bibfnamefont{K.}~\bibnamefont{Gaul}}, \bibinfo{author}{\bibfnamefont{R.}~\bibnamefont{Berger}}, \bibinfo{author}{\bibfnamefont{J.}~\bibnamefont{Billowes}}, \bibinfo{author}{\bibfnamefont{C.~L.} \bibnamefont{Binnersley}}, \bibinfo{author}{\bibfnamefont{M.~L.} \bibnamefont{Bissell}}, \bibinfo{author}{\bibfnamefont{A.~A.} \bibnamefont{Breier}}, \bibinfo{author}{\bibfnamefont{K.}~\bibnamefont{Chrysalidis}}, \bibnamefont{et~al.}, \bibinfo{journal}{Phys. Rev. Lett.} \textbf{\bibinfo{volume}{127}}, \bibinfo{pages}{033001} (\bibinfo{year}{2021}), \urlprefix\url{https://link.aps.org/doi/10.1103/PhysRevLett.127.033001}.

\bibitem[{\citenamefont{Pauling}(1931)}]{Pauling_paper}
\bibinfo{author}{\bibfnamefont{L.}~\bibnamefont{Pauling}}, \bibinfo{journal}{Journal of the American Chemical Society} \textbf{\bibinfo{volume}{53}}, \bibinfo{pages}{1367} (\bibinfo{year}{1931}), \urlprefix\url{https://doi.org/10.1021/ja01355a027}.

\bibitem[{\citenamefont{Pauling}(1986)}]{PaulingBook}
\bibinfo{author}{\bibfnamefont{L.}~\bibnamefont{Pauling}}, \emph{\bibinfo{title}{The nature of the chemical bond and the structure of molecules and crystals : an introduction to modern structural chemistry (3rd ed.)}} (\bibinfo{publisher}{Cornell University Press}, \bibinfo{address}{Ithaca, N.Y.}, \bibinfo{year}{1986}).

\bibitem[{\citenamefont{Hannay and Smyth}(1946)}]{Hannay1946}
\bibinfo{author}{\bibfnamefont{N.~B.} \bibnamefont{Hannay}} \bibnamefont{and} \bibinfo{author}{\bibfnamefont{C.~P.} \bibnamefont{Smyth}}, \bibinfo{journal}{Journal of the American Chemical Society} \textbf{\bibinfo{volume}{68}}, \bibinfo{pages}{171} (\bibinfo{year}{1946}), \eprint{https://doi.org/10.1021/ja01206a003}, \urlprefix\url{https://doi.org/10.1021/ja01206a003}.

\bibitem[{\citenamefont{Klessinger}(1970)}]{Klessinger1970}
\bibinfo{author}{\bibfnamefont{M.}~\bibnamefont{Klessinger}}, \bibinfo{journal}{Angewandte Chemie International Edition in English} \textbf{\bibinfo{volume}{9}}, \bibinfo{pages}{500} (\bibinfo{year}{1970}), \eprint{https://onlinelibrary.wiley.com/doi/pdf/10.1002/anie.197005001}, \urlprefix\url{https://onlinelibrary.wiley.com/doi/abs/10.1002/anie.197005001}.

\bibitem[{\citenamefont{\ifmmode~\acute{S}\else \'{S}\fi{}mia\l{}kowski and Tomza}(2021)}]{Tomza_Ag}
\bibinfo{author}{\bibfnamefont{M.}~\bibnamefont{\ifmmode~\acute{S}\else \'{S}\fi{}mia\l{}kowski}} \bibnamefont{and} \bibinfo{author}{\bibfnamefont{M.}~\bibnamefont{Tomza}}, \bibinfo{journal}{Phys. Rev. A} \textbf{\bibinfo{volume}{103}}, \bibinfo{pages}{022802} (\bibinfo{year}{2021}), \urlprefix\url{https://link.aps.org/doi/10.1103/PhysRevA.103.022802}.

\bibitem[{Note1()}]{Note1}
Note1, \bibinfo{note}{the main difference is in HF since Pauling assumed a 63~$\%$ of partial ionic character whereas Hanny and Smyth assumed a 41~$\%$.}

\bibitem[{\citenamefont{Coulson}(1952)}]{CoulsonBook}
\bibinfo{author}{\bibfnamefont{C.~A.} \bibnamefont{Coulson}}, \emph{\bibinfo{title}{Valence}} (\bibinfo{publisher}{Clarendon Press, Oxford}, \bibinfo{address}{Oxford, United Kingdom}, \bibinfo{year}{1952}).

\bibitem[{\citenamefont{Liu et~al.}(2025)\citenamefont{Liu, Aguirre, Kane, Kendrick, and Hemmerling}}]{hwwm-1mn7}
\bibinfo{author}{\bibfnamefont{L.-R.} \bibnamefont{Liu}}, \bibinfo{author}{\bibfnamefont{M.}~\bibnamefont{Aguirre}}, \bibinfo{author}{\bibfnamefont{S.~R.} \bibnamefont{Kane}}, \bibinfo{author}{\bibfnamefont{B.~K.} \bibnamefont{Kendrick}}, \bibnamefont{and} \bibinfo{author}{\bibfnamefont{B.}~\bibnamefont{Hemmerling}}, \bibinfo{journal}{Phys. Rev. A} \textbf{\bibinfo{volume}{111}}, \bibinfo{pages}{062810} (\bibinfo{year}{2025}), \urlprefix\url{https://link.aps.org/doi/10.1103/hwwm-1mn7}.

\bibitem[{\citenamefont{Zhang et~al.}(2015)\citenamefont{Zhang, Steimle, Cheng, and Stanton}}]{Zhang2015AuCl}
\bibinfo{author}{\bibfnamefont{R.}~\bibnamefont{Zhang}}, \bibinfo{author}{\bibfnamefont{T.~C.} \bibnamefont{Steimle}}, \bibinfo{author}{\bibfnamefont{L.}~\bibnamefont{Cheng}}, \bibnamefont{and} \bibinfo{author}{\bibfnamefont{J.~F.} \bibnamefont{Stanton}}, \bibinfo{journal}{Molecular Physics} \textbf{\bibinfo{volume}{113}}, \bibinfo{pages}{2073} (\bibinfo{year}{2015}), \eprint{https://doi.org/10.1080/00268976.2014.996619}, \urlprefix\url{https://doi.org/10.1080/00268976.2014.996619}.

\bibitem[{\citenamefont{Wang et~al.}(2009)\citenamefont{Wang, Zhuang, and Steimle}}]{10.1063/1.3226672}
\bibinfo{author}{\bibfnamefont{H.}~\bibnamefont{Wang}}, \bibinfo{author}{\bibfnamefont{X.}~\bibnamefont{Zhuang}}, \bibnamefont{and} \bibinfo{author}{\bibfnamefont{T.~C.} \bibnamefont{Steimle}}, \bibinfo{journal}{The Journal of Chemical Physics} \textbf{\bibinfo{volume}{131}}, \bibinfo{pages}{114315} (\bibinfo{year}{2009}), ISSN \bibinfo{issn}{0021-9606}, \eprint{https://pubs.aip.org/aip/jcp/article-pdf/doi/10.1063/1.3226672/15853028/114315\_1\_online.pdf}, \urlprefix\url{https://doi.org/10.1063/1.3226672}.

\bibitem[{\citenamefont{Zhuang and Steimle}(2014)}]{Zhuang2014CoO}
\bibinfo{author}{\bibfnamefont{X.}~\bibnamefont{Zhuang}} \bibnamefont{and} \bibinfo{author}{\bibfnamefont{T.~C.} \bibnamefont{Steimle}}, \bibinfo{journal}{The Journal of Chemical Physics} \textbf{\bibinfo{volume}{140}}, \bibinfo{pages}{124301} (\bibinfo{year}{2014}), \urlprefix\url{https://doi.org/10.1063/1.4868551}.

\bibitem[{\citenamefont{Koelemay and Ziurys}(2023)}]{Koelemay2023FeC}
\bibinfo{author}{\bibfnamefont{L.~A.} \bibnamefont{Koelemay}} \bibnamefont{and} \bibinfo{author}{\bibfnamefont{L.~M.} \bibnamefont{Ziurys}}, \bibinfo{journal}{The Astrophysical Journal Letters} \textbf{\bibinfo{volume}{958}}, \bibinfo{pages}{L6} (\bibinfo{year}{2023}), \urlprefix\url{https://doi.org/10.3847/2041-8213/ad0899}.

\bibitem[{\citenamefont{Steimle et~al.}(2006{\natexlab{a}})\citenamefont{Steimle, Chen, Harrison, and Brown}}]{Steimle2006FeH}
\bibinfo{author}{\bibfnamefont{T.~C.} \bibnamefont{Steimle}}, \bibinfo{author}{\bibfnamefont{J.}~\bibnamefont{Chen}}, \bibinfo{author}{\bibfnamefont{J.~J.} \bibnamefont{Harrison}}, \bibnamefont{and} \bibinfo{author}{\bibfnamefont{J.~M.} \bibnamefont{Brown}}, \bibinfo{journal}{The Journal of Chemical Physics} \textbf{\bibinfo{volume}{124}}, \bibinfo{pages}{184307} (\bibinfo{year}{2006}{\natexlab{a}}), \urlprefix\url{https://doi.org/10.1063/1.2194551}.

\bibitem[{\citenamefont{DeLeeuw and Dymanus}(1973)}]{DeLeeuw1973CoF}
\bibinfo{author}{\bibfnamefont{F.~A.} \bibnamefont{DeLeeuw}} \bibnamefont{and} \bibinfo{author}{\bibfnamefont{A.}~\bibnamefont{Dymanus}}, \bibinfo{journal}{Journal of Molecular Spectroscopy} \textbf{\bibinfo{volume}{48}}, \bibinfo{pages}{427} (\bibinfo{year}{1973}).

\bibitem[{\citenamefont{Le et~al.}(2013)\citenamefont{Le, Steimle, Skripnikov, and Titov}}]{10.1063/1.4794049}
\bibinfo{author}{\bibfnamefont{A.}~\bibnamefont{Le}}, \bibinfo{author}{\bibfnamefont{T.~C.} \bibnamefont{Steimle}}, \bibinfo{author}{\bibfnamefont{L.}~\bibnamefont{Skripnikov}}, \bibnamefont{and} \bibinfo{author}{\bibfnamefont{A.~V.} \bibnamefont{Titov}}, \bibinfo{journal}{The Journal of Chemical Physics} \textbf{\bibinfo{volume}{138}}, \bibinfo{pages}{124313} (\bibinfo{year}{2013}), ISSN \bibinfo{issn}{0021-9606}, \eprint{https://pubs.aip.org/aip/jcp/article-pdf/doi/10.1063/1.4794049/15462617/124313\_1\_online.pdf}, \urlprefix\url{https://doi.org/10.1063/1.4794049}.

\bibitem[{\citenamefont{Zhuang et~al.}(2010)\citenamefont{Zhuang, Steimle, and Linton}}]{10.1063/1.3505141}
\bibinfo{author}{\bibfnamefont{X.}~\bibnamefont{Zhuang}}, \bibinfo{author}{\bibfnamefont{T.~C.} \bibnamefont{Steimle}}, \bibnamefont{and} \bibinfo{author}{\bibfnamefont{C.}~\bibnamefont{Linton}}, \bibinfo{journal}{The Journal of Chemical Physics} \textbf{\bibinfo{volume}{133}}, \bibinfo{pages}{164310} (\bibinfo{year}{2010}), ISSN \bibinfo{issn}{0021-9606}, \eprint{https://pubs.aip.org/aip/jcp/article-pdf/doi/10.1063/1.3505141/13521406/164310\_1\_online.pdf}, \urlprefix\url{https://doi.org/10.1063/1.3505141}.

\bibitem[{\citenamefont{Wang et~al.}(2007)\citenamefont{Wang, Virgo, Chen, and Steimle}}]{10.1063/1.2778427}
\bibinfo{author}{\bibfnamefont{H.}~\bibnamefont{Wang}}, \bibinfo{author}{\bibfnamefont{W.~L.} \bibnamefont{Virgo}}, \bibinfo{author}{\bibfnamefont{J.}~\bibnamefont{Chen}}, \bibnamefont{and} \bibinfo{author}{\bibfnamefont{T.~C.} \bibnamefont{Steimle}}, \bibinfo{journal}{The Journal of Chemical Physics} \textbf{\bibinfo{volume}{127}}, \bibinfo{pages}{124302} (\bibinfo{year}{2007}), ISSN \bibinfo{issn}{0021-9606}, \eprint{https://pubs.aip.org/aip/jcp/article-pdf/doi/10.1063/1.2778427/10925232/124302\_1\_online.pdf}, \urlprefix\url{https://doi.org/10.1063/1.2778427}.

\bibitem[{\citenamefont{Fletcher et~al.}(1993)\citenamefont{Fletcher, Dai, Steimle, and Balasubramanian}}]{10.1063/1.465503}
\bibinfo{author}{\bibfnamefont{D.~A.} \bibnamefont{Fletcher}}, \bibinfo{author}{\bibfnamefont{D.}~\bibnamefont{Dai}}, \bibinfo{author}{\bibfnamefont{T.~C.} \bibnamefont{Steimle}}, \bibnamefont{and} \bibinfo{author}{\bibfnamefont{K.}~\bibnamefont{Balasubramanian}}, \bibinfo{journal}{The Journal of Chemical Physics} \textbf{\bibinfo{volume}{99}}, \bibinfo{pages}{9324} (\bibinfo{year}{1993}), ISSN \bibinfo{issn}{0021-9606}, \eprint{https://pubs.aip.org/aip/jcp/article-pdf/99/11/9324/19058597/9324\_1\_online.pdf}, \urlprefix\url{https://doi.org/10.1063/1.465503}.

\bibitem[{\citenamefont{Qin et~al.}(2012)\citenamefont{Qin, Zhang, Wang, and Steimle}}]{10.1063/1.4734596}
\bibinfo{author}{\bibfnamefont{C.}~\bibnamefont{Qin}}, \bibinfo{author}{\bibfnamefont{R.}~\bibnamefont{Zhang}}, \bibinfo{author}{\bibfnamefont{F.}~\bibnamefont{Wang}}, \bibnamefont{and} \bibinfo{author}{\bibfnamefont{T.~C.} \bibnamefont{Steimle}}, \bibinfo{journal}{The Journal of Chemical Physics} \textbf{\bibinfo{volume}{137}}, \bibinfo{pages}{054309} (\bibinfo{year}{2012}), ISSN \bibinfo{issn}{0021-9606}, \eprint{https://pubs.aip.org/aip/jcp/article-pdf/doi/10.1063/1.4734596/15452915/054309\_1\_online.pdf}, \urlprefix\url{https://doi.org/10.1063/1.4734596}.

\bibitem[{\citenamefont{Steimle et~al.}(1995)\citenamefont{Steimle, Jung, and Li}}]{10.1063/1.469750}
\bibinfo{author}{\bibfnamefont{T.~C.} \bibnamefont{Steimle}}, \bibinfo{author}{\bibfnamefont{K.~Y.} \bibnamefont{Jung}}, \bibnamefont{and} \bibinfo{author}{\bibfnamefont{B.}~\bibnamefont{Li}}, \bibinfo{journal}{The Journal of Chemical Physics} \textbf{\bibinfo{volume}{103}}, \bibinfo{pages}{1767} (\bibinfo{year}{1995}), ISSN \bibinfo{issn}{0021-9606}, \eprint{https://pubs.aip.org/aip/jcp/article-pdf/103/5/1767/19169676/1767\_1\_online.pdf}, \urlprefix\url{https://doi.org/10.1063/1.469750}.

\bibitem[{\citenamefont{Steimle and Virgo}(2004)}]{10.1063/1.1822917}
\bibinfo{author}{\bibfnamefont{T.~C.} \bibnamefont{Steimle}} \bibnamefont{and} \bibinfo{author}{\bibfnamefont{W.~L.} \bibnamefont{Virgo}}, \bibinfo{journal}{The Journal of Chemical Physics} \textbf{\bibinfo{volume}{121}}, \bibinfo{pages}{12411} (\bibinfo{year}{2004}), ISSN \bibinfo{issn}{0021-9606}, \eprint{https://pubs.aip.org/aip/jcp/article-pdf/121/24/12411/19158912/12411\_1\_online.pdf}, \urlprefix\url{https://doi.org/10.1063/1.1822917}.

\bibitem[{\citenamefont{Gengler et~al.}(2007)\citenamefont{Gengler, Ma, Adam, and Steimle}}]{10.1063/1.2711807}
\bibinfo{author}{\bibfnamefont{J.}~\bibnamefont{Gengler}}, \bibinfo{author}{\bibfnamefont{T.}~\bibnamefont{Ma}}, \bibinfo{author}{\bibfnamefont{A.~G.} \bibnamefont{Adam}}, \bibnamefont{and} \bibinfo{author}{\bibfnamefont{T.~C.} \bibnamefont{Steimle}}, \bibinfo{journal}{The Journal of Chemical Physics} \textbf{\bibinfo{volume}{126}}, \bibinfo{pages}{134304} (\bibinfo{year}{2007}), ISSN \bibinfo{issn}{0021-9606}, \eprint{https://pubs.aip.org/aip/jcp/article-pdf/doi/10.1063/1.2711807/15396366/134304\_1\_online.pdf}, \urlprefix\url{https://doi.org/10.1063/1.2711807}.

\bibitem[{\citenamefont{Steimle et~al.}(2006{\natexlab{b}})\citenamefont{Steimle, Virgo, and Ma}}]{10.1063/1.2145880}
\bibinfo{author}{\bibfnamefont{T.~C.} \bibnamefont{Steimle}}, \bibinfo{author}{\bibfnamefont{W.~L.} \bibnamefont{Virgo}}, \bibnamefont{and} \bibinfo{author}{\bibfnamefont{T.}~\bibnamefont{Ma}}, \bibinfo{journal}{The Journal of Chemical Physics} \textbf{\bibinfo{volume}{124}}, \bibinfo{pages}{024309} (\bibinfo{year}{2006}{\natexlab{b}}), ISSN \bibinfo{issn}{0021-9606}, \eprint{https://pubs.aip.org/aip/jcp/article-pdf/doi/10.1063/1.2145880/13217248/024309\_1\_online.pdf}, \urlprefix\url{https://doi.org/10.1063/1.2145880}.

\bibitem[{\citenamefont{Wang and Steimle}(2011)}]{10.1063/1.3595469}
\bibinfo{author}{\bibfnamefont{F.}~\bibnamefont{Wang}} \bibnamefont{and} \bibinfo{author}{\bibfnamefont{T.~C.} \bibnamefont{Steimle}}, \bibinfo{journal}{The Journal of Chemical Physics} \textbf{\bibinfo{volume}{134}}, \bibinfo{pages}{201106} (\bibinfo{year}{2011}), ISSN \bibinfo{issn}{0021-9606}, \eprint{https://pubs.aip.org/aip/jcp/article-pdf/doi/10.1063/1.3595469/13538519/201106\_1\_online.pdf}, \urlprefix\url{https://doi.org/10.1063/1.3595469}.

\bibitem[{\citenamefont{Ladjimi and Tomza}(2024)}]{Tomza-2024}
\bibinfo{author}{\bibfnamefont{H.}~\bibnamefont{Ladjimi}} \bibnamefont{and} \bibinfo{author}{\bibfnamefont{M.}~\bibnamefont{Tomza}}, \bibinfo{journal}{Phys. Rev. A} \textbf{\bibinfo{volume}{109}}, \bibinfo{pages}{052814} (\bibinfo{year}{2024}), \urlprefix\url{https://link.aps.org/doi/10.1103/PhysRevA.109.052814}.

\bibitem[{\citenamefont{Tomza}(2013)}]{Tomza-Cr}
\bibinfo{author}{\bibfnamefont{M.}~\bibnamefont{Tomza}}, \bibinfo{journal}{Phys. Rev. A} \textbf{\bibinfo{volume}{88}}, \bibinfo{pages}{012519} (\bibinfo{year}{2013}), \urlprefix\url{https://link.aps.org/doi/10.1103/PhysRevA.88.012519}.

\bibitem[{\citenamefont{Tomza}(2021)}]{Tomza_2021}
\bibinfo{author}{\bibfnamefont{M.}~\bibnamefont{Tomza}}, \bibinfo{journal}{New Journal of Physics} \textbf{\bibinfo{volume}{23}}, \bibinfo{pages}{085003} (\bibinfo{year}{2021}), \urlprefix\url{https://dx.doi.org/10.1088/1367-2630/ac1696}.

\bibitem[{\citenamefont{Liu et~al.}(2023)\citenamefont{Liu, McKemmish, and Pérez-Ríos}}]{LiuCCSDT}
\bibinfo{author}{\bibfnamefont{X.}~\bibnamefont{Liu}}, \bibinfo{author}{\bibfnamefont{L.}~\bibnamefont{McKemmish}}, \bibnamefont{and} \bibinfo{author}{\bibfnamefont{J.}~\bibnamefont{Pérez-Ríos}}, \bibinfo{journal}{Phys. Chem. Chem. Phys.} \textbf{\bibinfo{volume}{25}}, \bibinfo{pages}{4093} (\bibinfo{year}{2023}), \urlprefix\url{http://dx.doi.org/10.1039/D2CP05060A}.

\bibitem[{\citenamefont{Allen et~al.}(1993)\citenamefont{Allen, Capitani, Kolks, and Sproul}}]{ALLEN1993647}
\bibinfo{author}{\bibfnamefont{L.~C.} \bibnamefont{Allen}}, \bibinfo{author}{\bibfnamefont{J.~F.} \bibnamefont{Capitani}}, \bibinfo{author}{\bibfnamefont{G.~A.} \bibnamefont{Kolks}}, \bibnamefont{and} \bibinfo{author}{\bibfnamefont{G.~D.} \bibnamefont{Sproul}}, \bibinfo{journal}{Journal of Molecular Structure} \textbf{\bibinfo{volume}{300}}, \bibinfo{pages}{647} (\bibinfo{year}{1993}), ISSN \bibinfo{issn}{0022-2860}, \urlprefix\url{https://www.sciencedirect.com/science/article/pii/002228609387053C}.

\bibitem[{\citenamefont{Ketelaar}(1953)}]{Ketelaar1953ChemicalConstitution}
\bibinfo{author}{\bibfnamefont{J.~A.~A.} \bibnamefont{Ketelaar}}, \emph{\bibinfo{title}{Chemical Constitution: An Introduction to the Theory of the Chemical Bond}} (\bibinfo{publisher}{Elsevier Publishing Company}, \bibinfo{address}{Houston}, \bibinfo{year}{1953}).

\bibitem[{Note2()}]{Note2}
Note2, \bibinfo{note}{note that in the original Van Arkel-Ketelaar Triangle, compounds in the left corner are classified as metallic rather than van der Waals.}

\bibitem[{\citenamefont{Williams and Rasmussen}(2006)}]{williams2006gaussian}
\bibinfo{author}{\bibfnamefont{C.~K.} \bibnamefont{Williams}} \bibnamefont{and} \bibinfo{author}{\bibfnamefont{C.~E.} \bibnamefont{Rasmussen}}, \emph{\bibinfo{title}{Gaussian processes for machine learning}}, vol.~\bibinfo{volume}{2} (\bibinfo{publisher}{MIT press Cambridge, MA}, \bibinfo{year}{2006}).

\bibitem[{\citenamefont{Liu et~al.}(2020{\natexlab{a}})\citenamefont{Liu, Meijer, and Pérez-Ríos}}]{liu2020universality}
\bibinfo{author}{\bibfnamefont{X.}~\bibnamefont{Liu}}, \bibinfo{author}{\bibfnamefont{G.}~\bibnamefont{Meijer}}, \bibnamefont{and} \bibinfo{author}{\bibfnamefont{J.}~\bibnamefont{Pérez-Ríos}}, \emph{\bibinfo{title}{On the universality of spectroscopic constants of diatomic molecules}} (\bibinfo{year}{2020}{\natexlab{a}}), \eprint{2005.07913}.

\bibitem[{\citenamefont{Liu et~al.}(2020{\natexlab{b}})\citenamefont{Liu, Meijer, and Pérez-Ríos}}]{Liudipole}
\bibinfo{author}{\bibfnamefont{X.}~\bibnamefont{Liu}}, \bibinfo{author}{\bibfnamefont{G.}~\bibnamefont{Meijer}}, \bibnamefont{and} \bibinfo{author}{\bibfnamefont{J.}~\bibnamefont{Pérez-Ríos}}, \bibinfo{journal}{Phys. Chem. Chem. Phys.} \textbf{\bibinfo{volume}{22}}, \bibinfo{pages}{24191} (\bibinfo{year}{2020}{\natexlab{b}}), \urlprefix\url{http://dx.doi.org/10.1039/D0CP03810E}.

\bibitem[{\citenamefont{Cranmer}(2023)}]{cranmerInterpretableMachineLearning2023}
\bibinfo{author}{\bibfnamefont{M.}~\bibnamefont{Cranmer}}, \emph{\bibinfo{title}{Interpretable {Machine} {Learning} for {Science} with {PySR} and {SymbolicRegression}.jl}} (\bibinfo{year}{2023}), \bibinfo{note}{arXiv:2305.01582 [astro-ph, physics:physics]}, \urlprefix\url{http://arxiv.org/abs/2305.01582}.

\bibitem[{\citenamefont{Werner et~al.}(2008)\citenamefont{Werner, Knowles, Lindh, Manby, {Sch\"{u}tz} et~al.}}]{molpro}
\bibinfo{author}{\bibfnamefont{H.-J.} \bibnamefont{Werner}}, \bibinfo{author}{\bibfnamefont{P.~J.} \bibnamefont{Knowles}}, \bibinfo{author}{\bibfnamefont{R.}~\bibnamefont{Lindh}}, \bibinfo{author}{\bibfnamefont{F.~R.} \bibnamefont{Manby}}, \bibinfo{author}{\bibfnamefont{M.}~\bibnamefont{{Sch\"{u}tz}}}, \bibnamefont{et~al.}, \emph{\bibinfo{title}{Molpro, a package of ab initio programs}} (\bibinfo{year}{2008}), \bibinfo{note}{see http://www.molpro.net}.

\bibitem[{\citenamefont{Schwerdtfeger and Nagle}(2019)}]{Polarizabilities}
\bibinfo{author}{\bibfnamefont{P.}~\bibnamefont{Schwerdtfeger}} \bibnamefont{and} \bibinfo{author}{\bibfnamefont{J.~K.} \bibnamefont{Nagle}}, \bibinfo{journal}{Molecular Physics} \textbf{\bibinfo{volume}{117}}, \bibinfo{pages}{1200} (\bibinfo{year}{2019}), \eprint{https://doi.org/10.1080/00268976.2018.1535143}, \urlprefix\url{https://doi.org/10.1080/00268976.2018.1535143}.

\bibitem[{\citenamefont{Kramida et~al.}(2024)\citenamefont{Kramida, {Yu.~Ralchenko}, Reader, and {and NIST ASD Team}}}]{NIST_ASD}
\bibinfo{author}{\bibfnamefont{A.}~\bibnamefont{Kramida}}, \bibinfo{author}{\bibnamefont{{Yu.~Ralchenko}}}, \bibinfo{author}{\bibfnamefont{J.}~\bibnamefont{Reader}}, \bibnamefont{and} \bibinfo{author}{\bibnamefont{{and NIST ASD Team}}}, \bibinfo{howpublished}{{NIST Atomic Spectra Database (ver. 5.12), [Online]. Available: {\tt{https://physics.nist.gov/asd}} [2025, July 3]. National Institute of Standards and Technology, Gaithersburg, MD.}} (\bibinfo{year}{2024}).

\bibitem[{\citenamefont{Lu et~al.}(2021)\citenamefont{Lu, Tang, Fu, Liu, and Ning}}]{AuF}
\bibinfo{author}{\bibfnamefont{Y.}~\bibnamefont{Lu}}, \bibinfo{author}{\bibfnamefont{R.}~\bibnamefont{Tang}}, \bibinfo{author}{\bibfnamefont{X.}~\bibnamefont{Fu}}, \bibinfo{author}{\bibfnamefont{H.}~\bibnamefont{Liu}}, \bibnamefont{and} \bibinfo{author}{\bibfnamefont{C.}~\bibnamefont{Ning}}, \bibinfo{journal}{The Journal of Chemical Physics} \textbf{\bibinfo{volume}{154}}, \bibinfo{pages}{074303} (\bibinfo{year}{2021}), ISSN \bibinfo{issn}{0021-9606}, \eprint{https://pubs.aip.org/aip/jcp/article-pdf/doi/10.1063/5.0038560/15585847/074303\_1\_online.pdf}, \urlprefix\url{https://doi.org/10.1063/5.0038560}.

\bibitem[{\citenamefont{Zhang et~al.}(2017)\citenamefont{Zhang, Yu, Steimle, and Cheng}}]{AuS}
\bibinfo{author}{\bibfnamefont{R.}~\bibnamefont{Zhang}}, \bibinfo{author}{\bibfnamefont{Y.}~\bibnamefont{Yu}}, \bibinfo{author}{\bibfnamefont{T.~C.} \bibnamefont{Steimle}}, \bibnamefont{and} \bibinfo{author}{\bibfnamefont{L.}~\bibnamefont{Cheng}}, \bibinfo{journal}{The Journal of Chemical Physics} \textbf{\bibinfo{volume}{146}}, \bibinfo{pages}{064307} (\bibinfo{year}{2017}), ISSN \bibinfo{issn}{0021-9606}, \eprint{https://pubs.aip.org/aip/jcp/article-pdf/doi/10.1063/1.4975816/14776955/064307\_1\_online.pdf}, \urlprefix\url{https://doi.org/10.1063/1.4975816}.

\bibitem[{\citenamefont{Amano and Hirota}(1973)}]{SF}
\bibinfo{author}{\bibfnamefont{T.}~\bibnamefont{Amano}} \bibnamefont{and} \bibinfo{author}{\bibfnamefont{E.}~\bibnamefont{Hirota}}, \bibinfo{journal}{Journal of Molecular Spectroscopy} \textbf{\bibinfo{volume}{45}}, \bibinfo{pages}{417} (\bibinfo{year}{1973}), ISSN \bibinfo{issn}{0022-2852}, \urlprefix\url{https://www.sciencedirect.com/science/article/pii/0022285273902129}.

\bibitem[{\citenamefont{Byfleet et~al.}(1971)\citenamefont{Byfleet, Carrington, and Russell}}]{SeF}
\bibinfo{author}{\bibfnamefont{C.}~\bibnamefont{Byfleet}}, \bibinfo{author}{\bibfnamefont{A.}~\bibnamefont{Carrington}}, \bibnamefont{and} \bibinfo{author}{\bibfnamefont{D.}~\bibnamefont{Russell}}, \bibinfo{journal}{Molecular Physics} \textbf{\bibinfo{volume}{20}}, \bibinfo{pages}{271} (\bibinfo{year}{1971}), \eprint{https://doi.org/10.1080/00268977100100251}, \urlprefix\url{https://doi.org/10.1080/00268977100100251}.

\bibitem[{\citenamefont{Story and Hebert}(1976)}]{XI}
\bibinfo{author}{\bibfnamefont{J.}~\bibnamefont{Story}, \bibfnamefont{T.~L.}} \bibnamefont{and} \bibinfo{author}{\bibfnamefont{A.~J.} \bibnamefont{Hebert}}, \bibinfo{journal}{The Journal of Chemical Physics} \textbf{\bibinfo{volume}{64}}, \bibinfo{pages}{855} (\bibinfo{year}{1976}), ISSN \bibinfo{issn}{0021-9606}, \eprint{https://pubs.aip.org/aip/jcp/article-pdf/64/2/855/18898914/855\_1\_online.pdf}, \urlprefix\url{https://doi.org/10.1063/1.432235}.

\bibitem[{\citenamefont{Frisch et~al.}(2016)\citenamefont{Frisch, Trucks, Schlegel, Scuseria, Robb, Cheeseman, Scalmani, Barone, Petersson, Nakatsuji et~al.}}]{g16}
\bibinfo{author}{\bibfnamefont{M.~J.} \bibnamefont{Frisch}}, \bibinfo{author}{\bibfnamefont{G.~W.} \bibnamefont{Trucks}}, \bibinfo{author}{\bibfnamefont{H.~B.} \bibnamefont{Schlegel}}, \bibinfo{author}{\bibfnamefont{G.~E.} \bibnamefont{Scuseria}}, \bibinfo{author}{\bibfnamefont{M.~A.} \bibnamefont{Robb}}, \bibinfo{author}{\bibfnamefont{J.~R.} \bibnamefont{Cheeseman}}, \bibinfo{author}{\bibfnamefont{G.}~\bibnamefont{Scalmani}}, \bibinfo{author}{\bibfnamefont{V.}~\bibnamefont{Barone}}, \bibinfo{author}{\bibfnamefont{G.~A.} \bibnamefont{Petersson}}, \bibinfo{author}{\bibfnamefont{H.}~\bibnamefont{Nakatsuji}}, \bibnamefont{et~al.}, \emph{\bibinfo{title}{Gaussian 16 {R}evision {C}.01}} (\bibinfo{year}{2016}), \bibinfo{note}{{G}aussian {I}nc. {W}allingford {CT}}.

\bibitem[{\citenamefont{Chai and Head-Gordon}(2008)}]{wb97xd}
\bibinfo{author}{\bibfnamefont{J.-D.} \bibnamefont{Chai}} \bibnamefont{and} \bibinfo{author}{\bibfnamefont{M.}~\bibnamefont{Head-Gordon}}, \bibinfo{journal}{Phys. Chem. Chem. Phys.} \textbf{\bibinfo{volume}{10}}, \bibinfo{pages}{6615} (\bibinfo{year}{2008}).

\bibitem[{\citenamefont{Weigend et~al.}(2003)\citenamefont{Weigend, Furche, and Ahlrichs}}]{def2}
\bibinfo{author}{\bibfnamefont{F.}~\bibnamefont{Weigend}}, \bibinfo{author}{\bibfnamefont{F.}~\bibnamefont{Furche}}, \bibnamefont{and} \bibinfo{author}{\bibfnamefont{R.}~\bibnamefont{Ahlrichs}}, \bibinfo{journal}{The Journal of chemical physics} \textbf{\bibinfo{volume}{119}}, \bibinfo{pages}{12753} (\bibinfo{year}{2003}).

\bibitem[{\citenamefont{Pritchard et~al.}(2019)\citenamefont{Pritchard, Altarawy, Didier, Gibson, and Windus}}]{basis1}
\bibinfo{author}{\bibfnamefont{B.~P.} \bibnamefont{Pritchard}}, \bibinfo{author}{\bibfnamefont{D.}~\bibnamefont{Altarawy}}, \bibinfo{author}{\bibfnamefont{B.}~\bibnamefont{Didier}}, \bibinfo{author}{\bibfnamefont{T.~D.} \bibnamefont{Gibson}}, \bibnamefont{and} \bibinfo{author}{\bibfnamefont{T.~L.} \bibnamefont{Windus}}, \bibinfo{journal}{Journal of chemical information and modeling} \textbf{\bibinfo{volume}{59}}, \bibinfo{pages}{4814} (\bibinfo{year}{2019}).

\bibitem[{\citenamefont{Feller}(1996)}]{basis2}
\bibinfo{author}{\bibfnamefont{D.}~\bibnamefont{Feller}}, \bibinfo{journal}{Journal of computational chemistry} \textbf{\bibinfo{volume}{17}}, \bibinfo{pages}{1571} (\bibinfo{year}{1996}).

\bibitem[{\citenamefont{Schuchardt et~al.}(2007)\citenamefont{Schuchardt, Didier, Elsethagen, Sun, Gurumoorthi, Chase, Li, and Windus}}]{basis3}
\bibinfo{author}{\bibfnamefont{K.~L.} \bibnamefont{Schuchardt}}, \bibinfo{author}{\bibfnamefont{B.~T.} \bibnamefont{Didier}}, \bibinfo{author}{\bibfnamefont{T.}~\bibnamefont{Elsethagen}}, \bibinfo{author}{\bibfnamefont{L.}~\bibnamefont{Sun}}, \bibinfo{author}{\bibfnamefont{V.}~\bibnamefont{Gurumoorthi}}, \bibinfo{author}{\bibfnamefont{J.}~\bibnamefont{Chase}}, \bibinfo{author}{\bibfnamefont{J.}~\bibnamefont{Li}}, \bibnamefont{and} \bibinfo{author}{\bibfnamefont{T.~L.} \bibnamefont{Windus}}, \bibinfo{journal}{Journal of chemical information and modeling} \textbf{\bibinfo{volume}{47}}, \bibinfo{pages}{1045} (\bibinfo{year}{2007}).

\end{thebibliography}

\end{document}